\documentclass[final,3p,times]{elsarticle}
\usepackage{amssymb}
\usepackage[]{amsmath}
\usepackage{graphics}
\usepackage{graphicx}
\usepackage{amssymb}
\usepackage[]{amsmath}
\usepackage{amsthm}
\usepackage{setspace}
\usepackage{epsfig}
\usepackage{subfigure}
\usepackage{multicol}
\usepackage{multirow}
\usepackage{bm}
\usepackage{tikz}
\usetikzlibrary{positioning, calc}

\definecolor{lightyellow}{rgb}{1,1,0.6}
\definecolor{lightblue}{rgb}{0.8,0.85,1}
\definecolor{lightpurple}{rgb}{0.8,0.7,1}
\usepackage{colortbl}
\usepackage{xcolor}
\usepackage{booktabs}

\usepackage{color}
\usepackage{pifont,bm}
\usepackage{mathrsfs}
\usepackage{booktabs}
\usepackage{diagbox}
\usepackage{makecell}

\newtheorem{remark}{Remark}

\biboptions{numbers,sort&compress}
\usepackage[colorlinks,linkcolor=red,anchorcolor=blue,citecolor=green]{hyperref}

\usepackage{amsmath}
\usepackage{anysize}
\usepackage{pgfplots}
\usepackage{pgfplotstable}
\usepgfplotslibrary{groupplots}
\pgfplotsset{compat=1.18}
\marginsize{2cm}{2cm}{0cm}{1.5cm}
\selectfont

\journal {}

\begin{document}
\begin{frontmatter}

\title{ Gradient-enhanced PINN with residual unit for studying forward-inverse problems of variable coefficient equations}
\author{Hui-Juan Zhou$^{a}$}

 \author{Yong Chen$^{b,c}$\corref{cor1}}
\ead{ychen@sei.ecnu.edu.cn}
\cortext[cor1]{Corresponding author. }

\address[label]{School of Science, Shanghai Maritime University, Shanghai, 201306, P.R.China}

\address[label2]{School of Mathematical Sciences, Key Laboratory of MEA (Ministry of Education) and Shanghai Key Laboratory of PMMP, East China Normal University, Shanghai, 200241, China}

\address[label3]{College of Mathematics and Systems Science, Shandong University of Science and Technology, Qingdao, 266590, China}

\begin{abstract}

Physics-informed neural network (PINN) is a powerful emerging method for studying forward-inverse problems of partial differential equations (PDEs), even from limited sample data.
Variable coefficient PDEs, which model real-world phenomena, are of considerable physical significance and research value. This study proposes a gradient-enhanced PINN with residual unit (R-gPINN) method to solve the data-driven solution and function discovery for variable coefficient PDEs. On the one hand, the proposed method incorporates residual units into the neural networks to mitigate gradient vanishing and network degradation, unify linear and nonlinear coefficient problem. We present two types of residual unit structures in this work to offer more flexible solutions in problem-solving. On the other hand, by including gradient terms of variable coefficients, the method penalizes collocation points that fail to satisfy physical properties. This enhancement improves the network's adherence to physical constraints and aligns the prediction function more closely with the objective function. 
Numerical experiments including solve the forward-inverse problems of variable coefficient Burgers equation, variable coefficient KdV equation, variable coefficient Sine-Gordon equation, and high-dimensional variable coefficient Kadomtsev–Petviashvili equation. The results show that using R-gPINN method  can greatly improve the accuracy of predict solution and predict variable coefficient in solving variable coefficient equations. 

\end{abstract}

\end{frontmatter}

\section{Introduction}

The variable coefficient partial differential equations (PDEs) is receiving increasing attention due to its wide range of physical applications. Due to its coefficients are determined by specific physical quantities and need to be dynamically adjusted according to actual situations, the variable coefficients PDEs can describe a wider range of scientific fields than constant coefficient PDEs.  Variable coefficient PDEs have been widely used in fields of marine atmosphere, thermodynamics \cite{rlx2}, fluid mechanics \cite{ltlx1}, and quantum mechanics \cite{lzlx2}.  They can even be used to describe chemical reactions or biological diffusion processes \cite{hxsw2}. For example, the variable coefficient Korteweg-de Vries (KdV) equation has been developed to describe solitary waves in water with variable depth \cite{k-1971}, internal gravity waves in lakes with varying cross-sections \cite{G-1978}, and ion acoustic waves in inhomogeneous plasmas \cite{n-1975,k-1978}. The variable coefficients KdV equation can also model the damping of solitary waves in certain scenarios \cite{Ott-1970}. Due to the presence of arbitrary distribution functions in the solution of variable coefficient systems, different function forms can be selected to effectively regulate the solution of PDEs, thereby better explaining phenomena in practical applications. Variable coefficient systems often exhibit richer and more complex dynamic behaviors, and studying the dynamic behavior of variable coefficient systems can help us better understand practical problems.
Variable coefficient PDEs correspond to non spectral problems, methods for studying spectra cannot be directly used. Therefore, solving variable coefficient PDEs is still difficult compared to solving constant coefficient PDEs. Due to the wide physical applications of variable coefficient equations, solving the forward and inverse problems of variable coefficient equations is a very important and meaningful work.

Traditional numerical methods such as finite difference, finite element, and finite volume methods have been well-established in solving numerical solutions of PDEs, often yielding high-precision results. However, for the high-dimensional and inverse problems addressed in this study, Physics-Informed Neural Networks (PINNs) demonstrate unique advantages over traditional numerical methods.
The PINN method, as an emerging deep learning approach, was proposed by the Karniadakis team in 2019\cite{B3}. PINNs incorporate physical prior information, such as physical laws or strong form PDEs, directly into the loss function. This approach transforms the given PDE problem into an optimization problem involving network parameters. The core concept of PINNs is the universal approximation theorem\cite{Hornik-1989}, which allows for the accurate calculation of derivatives at sampling points using automatic differentiation techniques. By integrating strong form physical prior knowledge of PDE, the complexity of the hypothesis space is greatly reduced.
 The PINN method can high-precision predictions the solutions and unknown model parameters of PDE using only minimal sample data. Therefore, it is particularly suitable for solving high-dimensional problems and inverse problems.For high-dimensional equations, the PINN method, as a mesh-free approach, can mitigate the curse of dimensionality\cite{2024Karnidarks}. In addition, PINN demonstrates significant advantages in solving inverse problems. The model operates in a dual-driven mode by both data and physics, which is more effective than traditional numerical methods based on data assimilation. It can even address inverse problems that are challenging for traditional numerical methods to solve.
Due to the experimental data is often expensive and difficult to obtain, PINN offers significant advantages in the study of nonlinear systems and has led to a series of groundbreaking research achievements. A review article published in \textit{Nature Review Physics} in 2021, discusses the practical applications of PINN in areas such as turbulence simulation, fluid dynamics, heat transfer problems, three-dimensional wake flows, supersonic flows, biomedical engineering, nanophotonics, aerospace engineering, and materials science\cite{karniadakis-2021}.

In order to explore new methods for studying integrable systems, integrable systems theory combine with the PINN algorithm was proposed to address forward and inverse problems of integrable systems\cite{vcbugers7,Pu1-q,Pu1-lax,lin-miura,lin-jcp}. At present, the PINN method has achieved great success in solving the forward-inverse problems of constant coefficient integrable systems\cite{vcbugers8,vcbugers9,vcbugers10,vcbugers11,vcbugers12,vcbugers13}. For the variable coefficients equation, some researchers have attempted to use different modified PINN methods to solve the forward and inverse problems of the variable coefficient equation. In \cite{vcbugers48}, a PINNs method with two neural networks is proposed to solve the function discovery problem  under various noise levels of variable coefficient Hirota equation. In \cite{vcbugers49}, a parallel PINNs method with regularization term is proposed to investigate forward-inverse problems of the variable coefficient modified KdV system. More recently, variable coefficient PINN method\cite{Miao1} and gradient-enhanced PINN method based on transfer learning\cite{Lin1} have been developed to address inverse problems of variable coefficient systems. Notably, these methods require training two separate networks for the solution function and variable coefficient function, respectively. In \cite{yan-2023}, a modified PINNs method was proposed to study soliton dynamics and complex potentials recognition for 1D and 2D $\mathcal{P} \mathcal{T}$ symmetry saturable nonlinear Schr\"{o}dinger equations.
 
Although the traditional PINN method performs well for many PDE problems, its performance may be limited when handling variable coefficient PDEs. For instance, in the study of the inverse problem for variable coefficient equations in this paper, the input and output dimensions of the neural network are different. As a result, when using the traditional PINN method, the predicted values are often suboptimal. To solve the forward-inverse problems of variable coefficient equations, this study proposes a gradient-enhanced PINN with residual units (R-gPINN). This method combines the benefits of variable coefficient gradient-enhanced effects with the identity mapping ability of Residual Networks (ResNet)\cite{H-2016}. First, ResNet have the ability of identity mappings via adds shortcut connections between network layers. Introducing residual units into neural networks can effectively alleviate the problems of vanishing gradients and network degradation. For the function discovery problems of variable coefficient equations, the object function can be either linear or nonlinear, incorporates residual units into the neural networks can uniform linear and nonlinear coefficient problems effectively. We employ two types of residual unit structures in this work,differing in the placement of the activation function, providing more flexible solutions to problem-solving by selecting  different residual units and the number of hidden layers in the residual block.
Second, by incorporating the gradient term of the variable coefficient into the loss function of the R-gPINN, which penalizes collocation points that do not meet the physical properties of the variable coefficient. This approach enhances the network’s physical constraints, guiding the prediction function to progressively align with the objective function.
Numerical experiments  were conducted using the  variable coefficient Sine-Gordon equation, variable coefficient Burgers equation, variable coefficient KdV equation, variable coefficient Sine-Gordon (SG) equation, and high-dimensional variable coefficient Kadomtsev-Petviashvili (KP) equation. The results demonstrate that the R-gPINN method significantly improve the generalization ability of the network in the forward-inverse problems of the variable coefficient PDEs.

The structure of this article is as follows. In Section 2, a R-gPINN methods is proposed combined the advantages of variable coefficient gradient-enhanced effects and ResNet, for improve the performance of solving forward-inverse problems of variable coefficient equations. Section 3 applies the R-gPINN method to forward problems of the variable coefficient Burgers equation and compares its performance with the PINN method. Section 4 discusses the application of the R-gPINN method for identifying unknown variable coefficients in both linear and nonlinear equations. Numerical results underscore the R-PINN method's superior accuracy enhancement. Finally, the conclusions and discussions is given in the last Section.

\section{Methodology}

Consider the variable coefficient PDE with Dirichlet boundary conditions:
\begin{align}\label{E1}
\begin{cases}
f\left(\mathbf{x}, t ; \frac{\partial u}{\partial x_1}, \ldots, \frac{\partial u}{\partial x_N}, \frac{\partial u}{\partial t} ; \frac{\partial^2 u}{\partial x_1^2}, \ldots, \frac{\partial^2 u}{\partial x_1 \partial x_N}, \frac{\partial^2 u}{\partial x_1 \partial t} ; \ldots ; V(t)\right)=0, \\
u\left(\mathbf{x}, t_0\right)=u_0(\mathbf{x}), \ \ \mathbf{x} \in \Omega, \\
u(\mathbf{x}_{B}, t)=B(\mathbf{x}, t), \ \ \mathbf{x}_{B} \in \partial \Omega, \ \ t \in \left[t_0, t_1\right],
\end{cases}
\end{align}
where $\mathbf{x}=\left(x_1, \ldots, x_N\right) \in \Omega$,  $\partial \Omega$ is the boundary of  $\Omega$,$u=u(\mathbf{x}, t)$ is the solution function, $\Omega$ is a subset of $\mathbb{R}^N$, and $V(t)$ represents the variable coefficients in the PDE. 

\subsection{The fully connected feed-forward neural networks with residual unit}

The PINN method has demonstrated significant success in solving PDEs. Its network structure essentially evolves from fully connected neural networks. Consider constructing a fully connected feed-forward neural network with depth $d$ for solving PDEs. The network consists of one input layer, $d-1$ hidden layers, and one output layer. Let $N_i$ represent the number of neurons in the $i$-th hidden layer, where the output from the previous layer serves as the input for the next layer.

To introduce nonlinearity between neurons within each layer, a nonlinear activation function $\varphi$ is added to each neuron. The final mapping for the $i$-th layer can be expressed as:
\begin{equation}
	\mathbf{x}_i=\varphi(\mathbf{W}_i\mathbf{x}_{i-1}+\mathbf{b}_i),
\end{equation}
where $\mathbf{W}_i \in \mathbb{R}^{N_{i-1}\times N_i}$ and $\mathbf{b}_i \in \mathbb{R}^{N_i}$ represent the network weights matrix and bias vector of the $i$-th layer. Define an affine transformation as
\begin{equation}
	\mathcal{A}_i=\mathcal{A}_i(\mathbf{x}_{i-1})=\mathbf{W}_i\mathbf{x}_{i-1}+\mathbf{b}_i, \mathbf{x}_{i-1} \in \mathbb{R}^{N_{i-1}}, \mathbf{x}_i \in \mathbb{R}^{N_i}.
\end{equation}
With $\mathbf{x}^0 = \left(\mathbf{x}, t\right)$ as the input, the output $\mathbf{u}(\mathbf{x}^0; \bar{\Theta})$ of the fully connected feed-forward neural network can be represented as:
\begin{equation}\label{PINN_eq}
	\textbf{u}(\mathbf{x}^0;\Theta)=\big(\mathcal{A}_d\circ\varphi\circ \mathcal{A}_{d-1}\circ\cdots\circ\varphi\circ\mathcal{A}_1\big)(\mathbf{x}^0).
\end{equation}
Here, $\bar{\Theta} = \big\{\textbf{W}_i, \textbf{b}_i\big\}_{i=1}^{d}$ represents the trainable parameters of the network, and the output $\textbf{u}$ is used to approximate the solution of the problem.

In the context of variable coefficient problems, particularly for function discovery in variable coefficient equations, the coefficients may be either linear or nonlinear. Previous studies have generally used linear activation functions for linear targets and nonlinear activation functions for nonlinear targets to achieve higher precision. However, ResNet, by incorporating identity mappings, effectively addresses both linear and nonlinear problems uniformly. In this study, we consistently use the tanh function as the activation function to handle various types of problems. We introduce shortcut connections into the fully connected feed-forward neural network.

Assuming each residual block contains \( n \) hidden layers, we utilize two different types of residual blocks in this study. For clarity, the structure of the \( m \)-th residual block is illustrated below:
\begin{figure}[htbp]
\subfigure[]{\label{fig_pre_act_res_unit}
\begin{minipage}[t]{0.5\textwidth}
\centering
\includegraphics[width=\textwidth]{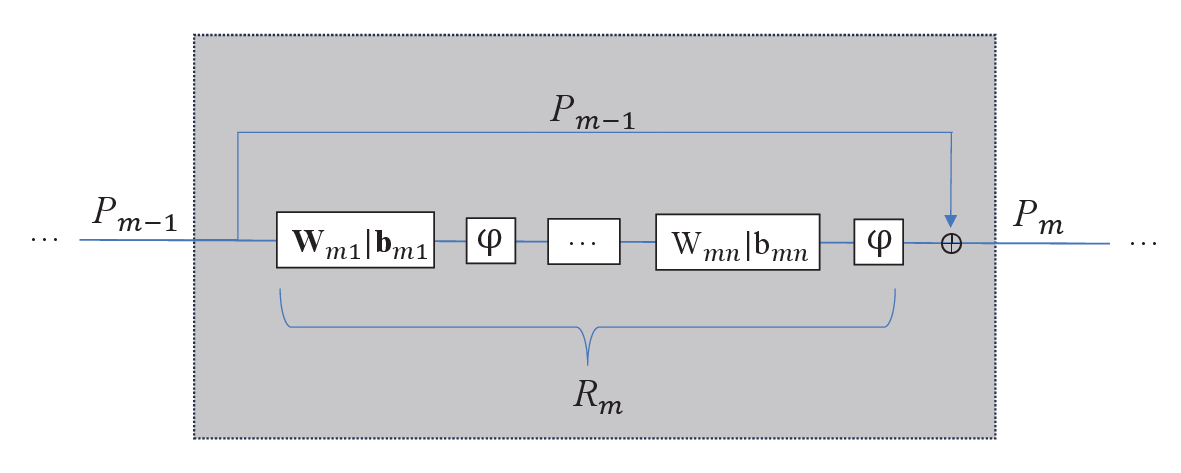}
\end{minipage}}%
\subfigure[]{\label{fig_out_act_res_unit}
\begin{minipage}[t]{0.5\textwidth}
\centering
\includegraphics[width=\textwidth]{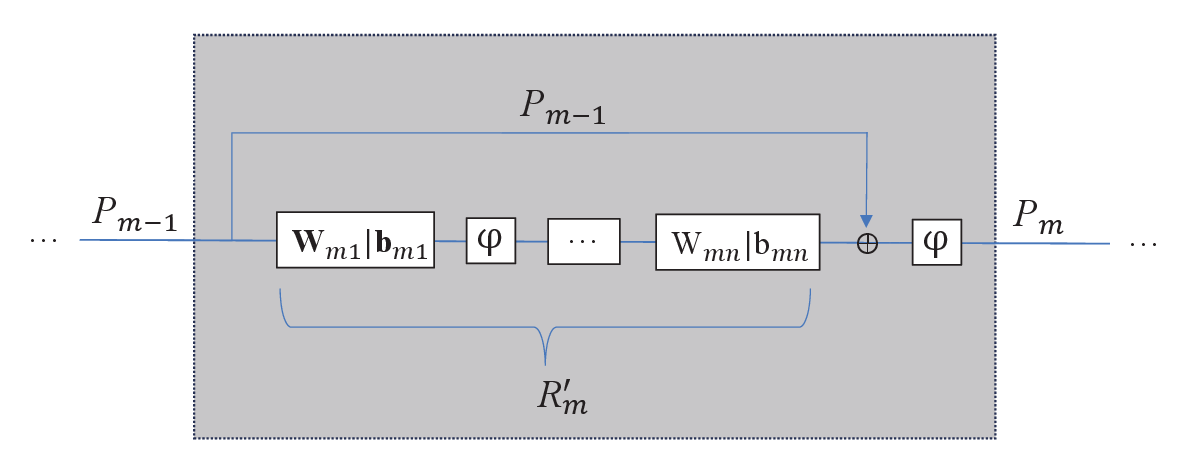}
\end{minipage}%
}%
\centering
\caption{ (Color online) (a) Pre-activation residual unit. (b) Post-activation residual unit.}
\label{Residual Unit}
\end{figure}

where \(P_{m-1}\) and \(P_{m}\) are the input and output of the \(m\)-th residual unit, and \(R_{m}\) and \(R^{'}_{m}\) are the residual functions for different types of residual units. 
$$R_{m}=R_{m}(P_{m-1})=\big(\varphi\circ\mathcal{A}_{mn}\circ\varphi\circ \mathcal{A}_{mn-1}\circ\cdots\circ\varphi\circ\mathcal{A}_{m1}\big)(P_{m-1}),$$ 
$$R^{'}_{m}=R^{'}_{m}(P_{m-1})=\big(\mathcal{A}_{mn}\circ\varphi\circ \mathcal{A}_{mn-1}\circ\cdots\circ\varphi\circ\mathcal{A}_{m1}\big)(P_{m-1}).$$ 

The two residual connections differ in the placement of the activation function: the pre-activation Residual unit calculates the activation function before the residual connection, while the post-activation residual unit performs the activation function calculation after the residual connection. In the pre-activation residual unit (see Figure \ref{fig_pre_act_res_unit}), the activation function is applied before the residual connection. Applying an activation function to each layer helps ensure that gradients do not disappear or explode during backpropagation, simplifying gradient propagation. This is particularly important for deep networks. Additionally, applying activation functions at each layer introduces more nonlinearities, enhancing the model's expressiveness. In the post-activation residual unit (see Figure \ref{fig_out_act_res_unit}), the residual connection is followed by an activation function calculation. Applying the activation function after the residual connection smooths the residual connection and reduces nonlinear noise, making the network more stable and easier to train, thus improving computational efficiency, especially in shallow networks.

The pre-activation residual connection \(P_{m} = P_{m-1} + R_{m}\) is executed only when the dimensions of \(P_{m-1}\) and \(R_{m}\) match. If this condition is not met, the residual connection is skipped, and the program proceeds without adding the previous layer's output to the current layer's output. Through residual connections, if a certain residual block fails to learn useful features (i.e., the residual function becomes zero), the network will directly pass the input to the next layer via the skip connection, achieving an identity mapping. This means that if a certain layer does not need to perform any transformation, the network can learn an identity mapping, making the output equal to the input, thereby preventing performance degradation.

In this study, we assume each hidden layer has \(N_d\) neurons. We did not use linear projection to match the dimensions of the input layer and the first hidden layer, so the first residual unit does not actually include a shortcut connection:
$$P_{1}(P_{0})=P_{1}=R_{1}=R_{1}(P_{0})=\big(\varphi\circ\mathcal{A}_{n}\circ\varphi\circ \mathcal{A}_{n-1}\circ\cdots\circ\varphi\circ\mathcal{A}_{1}\big)(P_{0}),$$ 
$$P_{0}=\mathbf{x}^0.$$

For the post-activation residual unit \(P_{m} = \varphi(P_{m-1} + R^{'}_{m})\), the situation is similar:
$$P_{1}(P_{0})=P_{1}=R^{'}_{1}=R^{'}_{1}(P_{0})=\big(\mathcal{A}_{n}\circ\varphi\circ \mathcal{A}_{n-1}\circ\cdots\circ\varphi\circ\mathcal{A}_{1}\big)(P_{0}).$$

Additionally, when the total number of hidden layers \(d-1\) is not divisible by the number \(n\) of hidden layers in the residual block, the shortcut connection will automatically be removed from the remaining hidden layers. The output \(\mathbf{u}(\mathbf{x}^0; \bar{\Theta})\) of the fully connected feed-forward neural network with residual connections can be represented as:
\begin{equation}
    \mathbf{u}(\mathbf{x}^0; \bar{\Theta}) = \left(\mathcal{A}_{d}\circ \varphi\circ \mathcal{A}_{d-1}\circ \cdots  \circ \varphi\circ\mathcal{A}_{[\frac{d-1}{n}]n+1} \circ P_{[\frac{d-1}{n}]} \circ \cdots \circ P_{2}\circ P_{1}\right)(\mathbf{x}^0),
\end{equation}

\subsection{Loss function for the forward-fnverse problems}

Define
\begin{equation}\label{PI}
f_u := f\left(\mathbf{x}, t ; \frac{\partial u}{\partial x_1}, \ldots, \frac{\partial u}{\partial x_N}, \frac{\partial u}{\partial t} ; \frac{\partial^2 u}{\partial x_1^2}, \ldots, \frac{\partial^2 u}{\partial x_1 \partial x_N}, \frac{\partial^2 u}{\partial x_1 \partial t} ; \ldots ; V(t)\right),
\end{equation}
where the fully connected feed-forward neural networks with residual unit, along with the physical governing equation \eqref{PI}, results in a physics-informed neural network with residual unit (R-PINN). The shared parameters between the neural network and the physical information are learned by minimizing the mean squared error loss.

{\bf Loss function for forward problems}

The mean squared error loss function which used to measure the difference between predicted and true values, is defined as:
\begin{equation}\label{loss_f}
Loss_{F}=Loss_u+Loss_{f_u},
\end{equation}
where
\begin{equation}
	\begin{aligned}
		Loss_u &=\frac{1}{N_u} \sum\limits_{i=1}^{N_u}\left((\hat{u}(\textbf{x}_i,t_i)-u(\textbf{x}_i,t_i))^2\right), \\
		Loss_{f_u} &= \frac{1}{N_{f_u}}  \sum\limits_{j=1}^{N_{f_u}}\left((\hat{f}_u(\textbf{x}_j,t_j))^2\right).
	\end{aligned}
\end{equation}
Here, $\{\textbf{x}_i, t_i, u(\textbf{x}_i, t_i)\}_{i=1}^{N_u}$ represent the initial and boundary value data for equation \eqref{E1}, and $N_u$ is the number of selected points. $\hat{u}(\textbf{x}_i, t_i)$ and $\hat{f}_u$ denote the predicted output values through PINN. $u(\textbf{x}_i, t_i)$ is the exact value of equation \eqref{E1} at the point $(\textbf{x}_i, t_i)$. Additionally, $\{\textbf{x}_j, t_j\}_{j=1}^{N_{f_u}}$ represent the collocation points of $f_u$ obtained by Latin Hypercube Sampling (LHS), $N_{f_u}$ is the number of collocation points, and $\hat{f}_u(\textbf{x}_j, t_j)$ is the predicted output of the physical governing equation by PINN.

{\bf Loss function for inverse problems}

For function discovery problems, where the variable coefficients $V(t)$ in equation \eqref{E1} are unknown, additional solution data $u_{in}$ is often required to learn the unknown $V(t)$ along with the solution $u$. Therefore, the loss function for the inverse problem is redefined as:

 We use a fully connected neural network \eqref{PINN_eq} with the Xavier initialization method. The activation function used in all examples is the hyperbolic tangent function, denoted as $\tanh(x)$. With $\mathbf{x}$ and $t$ as inputs, the network outputs $\hat{u}(\mathbf{x}, t)$ and $\hat{V}(\mathbf{x}, t)$. The variable coefficient in equation \eqref{E1} is independent of the spatial variables $\mathbf{x}$, meaning $V(\mathbf{x}, t) = V(t)$. Hence,
\begin{equation}\label{vcg}
\frac{\partial V(\mathbf{x}, t)}{\partial \mathbf{x}} = 0.
\end{equation}

Define
\begin{equation}\label{V_g}
V_g := \frac{\partial V(\mathbf{x}, t)}{\partial \mathbf{x}}.
\end{equation}

To solve the inverse problem of variable coefficient PDEs, we focus on the information of the variable coefficient. The gradient term of the variable coefficient can be added to the loss function  to enhance the constraint of physical information on the network\cite{gPINN}. This method is referred to as the gPINN  in this study, and its loss function is defined as:
\begin{equation}
	Loss_{I} = Loss_{u} + Loss_{f_u} + Loss_{u_{in}} + Loss_{V} + Loss_{V_g},
\end{equation}
where  $loss_u$, $loss_{f_u}$, $loss_V$, $loss_{V_g}$ are defined as follows:
\begin{equation}
	\begin{aligned}
	Loss_u &=\frac{1}{N_u} \sum\limits_{i=1}^{N_u}(\hat{u}(\textbf{x}_i,t_i)-u(\textbf{x}_i,t_i))^2, \\
	Loss_{f_u} &= \frac{1}{N_{f_u}}  \sum\limits_{j=1}^{N_{f_u}}(\hat{f}(\textbf{x}_j,t_j))^2,\\
	Loss_{u_{in}}&=\frac{1}{N_{u_{in}}}\sum^{N_{u_{in}}}_{j=1}\left|\hat{u}(\textbf{x}_{in}^j, t_{in}^j) - u(\textbf{x}_{in}^j, t_{in}^j)\right|^{2}, \\
		Loss_V &=\frac{1}{N_V} \sum\limits_{j=1}^{N_V}(\hat{V}(\textbf{x}_j,t_j)-V(t_j))^2, \\
		Loss_{V_g} &= \frac{1}{N_{V}}  \sum\limits_{j=1}^{N_{V}}(\frac{\delta\hat{V}(\textbf{x}_j,t_j)}{\delta \textbf{x}})^2.
	\end{aligned}
\end{equation}
Here, $\{\textbf{x}_{in}^j, t_{in}^j, u(\textbf{x}_{in}^j, t_{in}^j)\}_{j=1}^{N_{u_{in}}}$ represents the internal grid points dataset. $\hat{u}(\textbf{x}_{in}^j, t_{in}^j)$ and $\hat{V}(\textbf{x}_j,t_j)$ denote the predicted values. $\{t_{j}, V(t_{j})\}_{j=1}^{N_{V}}$ represents the training dataset of the variable coefficient \(V(t)\) at the initial and end of the time domain \([t_0, t_1]\).  The gradient term $Loss_{V_g}$ in the spatial direction is added to the loss function to ensure that the predicted variable coefficient $\hat{V}(\textbf{x}_j,t_j)$ is independent of spatial $\mathbf{x}$. $Loss_{V_g}$ penalizes collocation points that do not satisfy \eqref{V_g}, guiding the predicted variable coefficient $\hat{V}(\textbf{x}_j, t_j)$ to approximate the real  variable coefficient $V(t_j)$ during network training, thereby enhancing the network's generalization ability on the test set. In this work, the Adam method and the L-BFGS method are used to minimize the loss function by optimizing the parameters $\bar{\Theta}$ of the neural networks.

\begin{figure}[htbp]
    \centering
    \includegraphics[width=\textwidth]{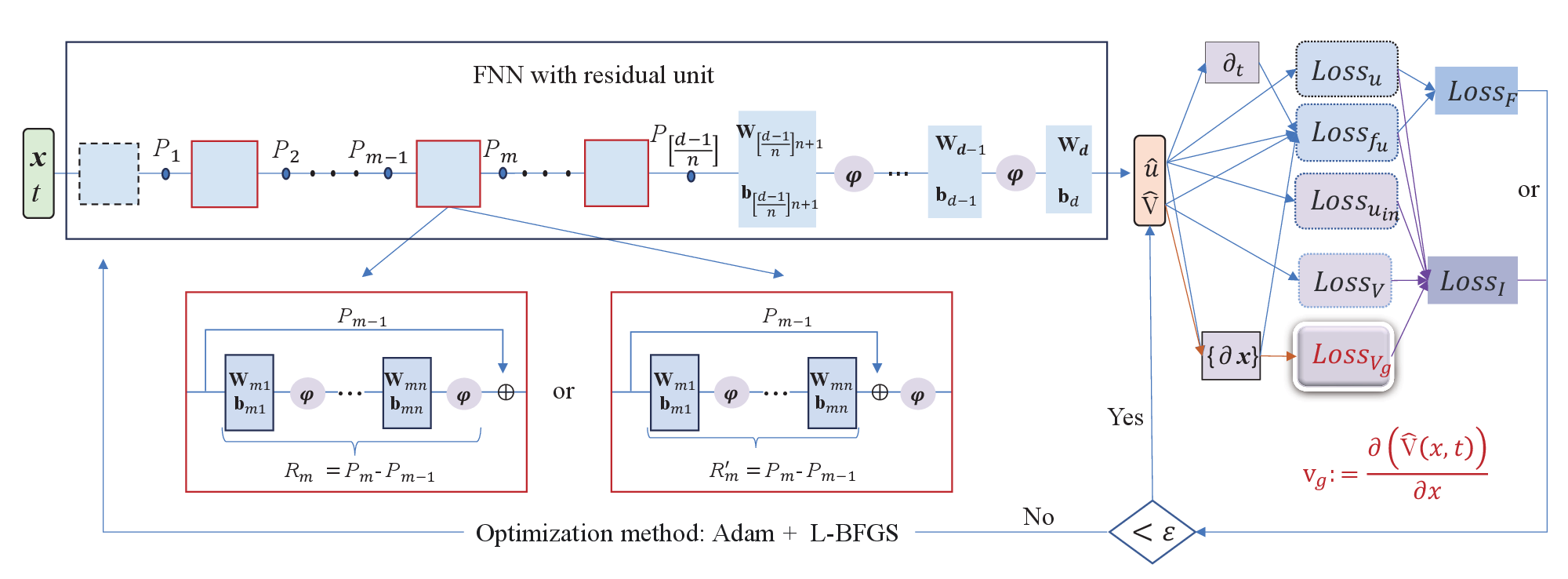}
    \caption{ (Color online) The flowchart of the R-gPINN algorithm for solving forward-inverse problems of variable coefficient PDEs: $\{\partial \mathbf{x}\}$ denotes the gradients with respect to $x$ required during the network training process.}
    \label{fig_R-gPINN}
\end{figure}

By incorporating residual unit  and gradient term of the variable coefficient into the PINN, the network's ability to handle both linear and nonlinear problems, and stabilize the training process of deep networks are improved. We name this method R-gPINN Method, leveraging the combined advantages of variable coefficient gradient-enhanced effects and ResNet. The flowchart of the R-gPINN algorithm for solving forward-inverse problems of variable coefficient PDEs is shown in Figure \ref{fig_R-gPINN}.

In the R-gPINN algorithm, we use a fully connected neural network with residual connections, with $\mathbf{x}$ and $t$ as the network inputs. The outputs of the network are $\hat{u}(\mathbf{x}, t)$ and $\hat{V}(\mathbf{x}, t)$. 

\subsection{Evaluation of R-gPINN's Generalization Ability}

To evaluate the performance of this methods, we will calculate the relative $\mathbb{L}_2$ error of the exact values and predicted values in the train domain. For 1+1 PDEs, assume $\Omega=\left[l_0, l_1\right]$ and $t \in \left[t_0, t_1\right]$. The spatial region $\left[l_0, l_1\right]$ and time region $\left[t_0, t_1\right]$ are discretized into $N_x$ and $N_t$ equidistant grid points, respectively. As a result, the solution $u$ is discretized into $N_x \times N_t$ data points within the given spatiotemporal domain. To obtain initial-boundary data on the aforementioned grids, we randomly select $N_u$ points denoted as $\left[l_0+j \frac{l_1-l_0}{N_x-1}, t_0\right] \cup \left[x, t_0+k \frac{t_1-t_0}{N_t-1}\right], j=0,1, \cdots, N_x-1, x=l_0$ or $l_1, k=0,1, \cdots, N_t-1$. Additionally, we randomly select $N_f$ collocation points for $f(x, t)$ in $\left[l_0, l_1\right] \times\left[t_0, t_1\right]$, which are not necessarily aligned with the grids. Thus, the training data in this case can be represented as $\left\{x_u^i, t_u^i, u^i\right\}_{i=1}^{N_u}$ and $\left\{x_f^i, t_f^i\right\}_{i=1}^{N_f}$. Since the size of the training data constitutes only a small percentage of the total grid data, the generalization ability of the R-gPINNs model is evaluated by calculating the relative $\mathbb{L}_2$ error  of the predicted solutions over the $N_x \times N_t$ data points on the grids:
$$
\text{error}_{u}=\frac{\sqrt{\sum_{j=0}^{N_x-1} \sum_{k=0}^{N_t-1}\left|\hat{u}\left(l_0+j \frac{l_1-l_0}{N_x-1}, t_0+k \frac{t_1-t_0}{N_{t}-1}\right)-u^{j, k}\right|^2}}{\sqrt{\sum_{j=0}^{N_x-1} \sum_{k=0}^{N_{t}-1}\left|u^{j, k}\right|^2}},
$$
where $\hat{u}\left(l_0+j \frac{l_1-l_0}{N_x-1}, t_0+k \frac{t_1-t_0}{N_t-1}\right)$ and $u^{j, k}$ represent the predicted solutions and exact solutions, respectively.

To calculate the error of the unknown function $V(t)$, the relative $\mathbb{L}_2$ norm error is defined as:
$$
\text{error}_{V}=\frac{\sqrt{\sum_{j=0}^{N_{V_x}-1} \sum_{k=0}^{N_{V_t}-1}\left|\hat{V}\left(l_0+j \frac{l_1-l_0}{N_{V_x}-1}, t_0+k \frac{t_1-t_0}{N_{V_t}-1}\right)-V^{j, k}\right|^2}}{\sqrt{\sum_{j=0}^{N_{V_x}-1} \sum_{k=0}^{N_{V_t}-1}\left|V^{j, k}\right|^2}},
$$
where $\hat{V}\left(l_0+j \frac{l_1-l_0}{N_{V_x}-1}, t_0+k \frac{t_1-t_0}{N_{V_t}-1}\right)$ and $V^{j, k}$ represent the predicted value and true value, respectively.  $N_{V_x}$ and $N_{V_t}$ denote  the equidistant points after dividing   the spatial region $\left[l_0, l_1\right]$ and time region $\left[t_0, t_1\right]$ into equal parts.

On the basis of traditional PINN, we will refer to the method of adding only variable coefficient gradient terms as g-PINN, and the method of adding only residual units in the network as R-PINN.
 For intuitive comparison of the prediction accuracy of different methods, the error reduction rate (ERR) can be directly calculated according to the relative $\mathbb{L}_2$ error:
\begin{equation}
\begin{aligned}
	\text{ERR} &= \frac{\text{error}^{\text{PINN}}-\text{error}^{\text{R-gPINN}}}{\text{error}^{\text{PINN}}}, \\
\text{ERR}_{13} &= \frac{\text{error}^{\text{R-PINN}}-\text{error}^{\text{R-gPINN}}}{\text{error}^{\text{R-PINN}}}, \\
\text{ERR}_{23} &= \frac{\text{error}^{\text{gPINN}}-\text{error}^{\text{R-gPINN}}}{\text{error}^{\text{gPINN}}}.
\end{aligned}
\end{equation}

\begin{remark}
The R-gPINN proposed in this work will apply to solve forward-inverse problems of variable coefficient PDE equation. In all examples, all the code in this study is based on Python 3.7 and Tensorflow 1.15 and implemented by using the DELL Precision 7920 Tower computer with a 2.10 GHz 8-core Xeon Silver 4110 processor, 64 GB memory and an 11 GB NVIDIA GeForce GTX 1080 Ti video card.
\end{remark}

\section{Results-Forward problem of variable coefficient PDE}

In addressing the data-driven forward problem of variable coefficient PDEs, where the variable coefficient terms are known, it is often sufficient to rely on the initial boundary value information of the solution. Consequently, the R-PINN method is well-suited for solving forward problems of variable coefficient PDEs. Given that the predicted solution $u$ is also available when tackling the inverse problem of variable coefficient PDEs, this chapter will focus on using the variable coefficient Burgers equation as an example to compare the performance of the R-PINN method against the traditional PINN method. The soliton solution for the variable coefficient Burgers equation, derived using a bilinear approach, is provided below.

\subsection{The soliton of the variable coefficient Burgers equation}

The variable coefficient Burgers equation is a fundamental equation in integrable systems, widely used to model the propagation and reflection of shock waves. It finds applications in various fields, such as fluid dynamics, nonlinear acoustics, and gas dynamics. The study of solutions to the variable coefficient Burgers equation dates back to the 1950s when Cole introduced the renowned Cole-Hopf transformation method while exploring quasi-linear parabolic equations in aerodynamics \cite{B18}. This method was instrumental in obtaining exact solutions to the variable coefficient Burgers equation. In the 1990s, Andrew Parker extended this research by deriving periodic solutions for the equation using a periodic method independent of the Cole-Hopf transformation \cite{B19}. Since the onset of the 21st century, research on the variable coefficient Burgers equation has entered a new phase, marked by the development of various novel methods and results. These include the Adomian decomposition method \cite{Ze1}, the variational iteration method \cite{Mo1}, the finite difference method \cite{Fl1}, and the finite element method \cite{Ali1}, among others. 

 The variable coefficient Burgers equation is expressed as follows:
\begin{equation}
	u_{t}+\alpha(t)uu_{x}+\beta(t)u_{xx} = 0, \label{eq_3.1}
\end{equation}
then we will use simplified bilinear method\cite{Al1} to solve the solutions of this equation. 

Substituting $f(x,t)=e^{kx-q(t)}= e^{\theta}$ into the linear term of \eqref{eq_3.1}, we obtain the dispersion relation:
 \begin{equation}
	q(t)=k^{2}\int \beta(t)dt,
	\label{eq_3.2}
\end{equation}
i.e.:
\begin{equation}
	\theta=kx-k^{2}\int \beta(t)dt.
	\label{eq_3.3}
\end{equation}
For equation \eqref{eq_3.1}, set soliton solution has the form as following:
 \begin{equation}
	u(x,t)=R(x,t)ln(1+e^{\theta})_{x}.
\end{equation}
By substituting \( u(x,t) \) into \eqref{eq_3.1} and assuming \( \beta(t) = m\alpha(t) \), we obtain:
\begin{equation}
	\begin{split}
		&kR_{t}+k^{2}\alpha(t)RR_{x}+km\alpha(t)R_{xx}=0,\\
		&k^{3}\alpha(t)R^{2}-k^{3}m\alpha(t)R-kRq'(t)+2kR_{t}+2k^{2}m\alpha(t)R_{x}+k^{2}\alpha(t)RR_{x}+2km\alpha(t)R_{xx}=0,\\
		&k^{3}m\alpha(t)R-kRq'(t)+kR_{t}+2k^{2}m\alpha(t)R_{x}+km\alpha(t)R_{xx}=0.  \label{eq_3.5}
	\end{split}
\end{equation}
Then  $R(x,t)=2m$ can be reduced, and the soliton solution of equation \eqref{eq_3.1} as follows:
 \begin{equation}
	u(x,t)=2m\frac{k_{1}e^{k_{1}x-k_{1}^{2}\int\beta(t)dt}}{1+e^{k_{1}x-k_{1}^{2}\int\beta(t)dt}}.
	\label{eq_3.7}
\end{equation}

\subsection{Data-driven solution of the variable coefficient Burgers equation}

When take $\beta(t)=\sin(t)$ and $k_{1}=m=1$ in \eqref{eq_3.7}, the variable coefficient Burgers equation has the exact soliton solution as below:
\begin{equation}\label{vcbuger-soliton}
	u_{b1}(x,t)=2\frac{e^{x+cos(t)}}{1+e^{x+cos(t)}}.
\end{equation}
For variable coefficient Burgers equation, the input $\left(\mathbf{x}, t\right)$=$(x,t)$,  now we take $\Omega$=$[-3,3]$,
$\left[t_0, t_1\right]$=$[-5,5]$ as the training region.
Consider the  variable coefficient Burgers equation with initial and boundary conditions:
\begin{equation}
\begin{cases}
		u_{t}+\sin(t)uu_{x}+\sin(t)u_{xx} = 0,  \\
		u(x,t_0) = u_{b1}(x,-5), \\
		u(x_l,t) = u_{b1}(-3,t), u(x_r,t)= u_{b1}(3,t),  (x,t)  \in [-3,3] \times [-5,5].
		\label{eq_4.1}
	\end{cases}
\end{equation}
Now, the spatial region $\left[-3, 3\right]$ and time region $\left[-5, 5\right]$ are discretized into $N_x=1001$ and $N_t=1001$ equidistant grid points, respectively. As a result, the solution $u$ is discretized into $1001 \times 1001$ data points within the given spatiotemporal domain. To obtain initial-boundary data on the aforementioned grids, we randomly select $N_u$ points denoted as $\left[-3+j \frac{6}{1000}, -5\right] \cup \left[x, -5+k \frac{10}{1000}\right], j=0,1, \cdots, 1000, x=-3$ or 3, $k=0,1, \cdots, 1000$. Additionally, we use LHS method select $N_f$ collocation points for $f(x, t)$ from region  $\left[-3, 3\right] \times\left[-5, 5\right]$, which are not necessarily aligned with the grids. Thus, the training dataset in this case can be represented as $\left\{x_u^i, t_u^i, u^i\right\}_{i=1}^{N_u}$ and $\left\{x_f^i, t_f^i\right\}_{i=1}^{N_f}$.

The physics information term in the R-PINN method is defined as:
\begin{equation}
f=u_{t}+\sin(t)uu_{x}+\sin(t)u_{xx}.
\end{equation}
The loss function for solving the variable coefficient Burgers equation using R-PINN method can be defined as:
\begin{equation}
\begin{aligned}
	Loss&=Loss_u+Loss_{f_u},\\
Loss_u &=\frac{1}{N_u} \sum\limits_{i=1}^{N_u}((\hat{u}(x_i,t_i)-u(x_i,t_i))^2), \\
		Loss_{f_u} &= \frac{1}{N_{f_u}}  \sum\limits_{j=1}^{N_{f_u}}((\hat{f}(x_j,t_j))^2).
	\end{aligned}
\end{equation}
Where $u(x_i,t_i)$ and $\hat{u}(x_i,t_i)$ represent the exact boundary value and predict value corresponding to the random selected points $(x_i,t_i)$ of the variable coefficient Burgers equation \eqref{eq_3.1} with  $\alpha(t)=\beta(t)=\sin(t)$.  $\hat{f}(x_j,t_j)$ represents the predict value corresponding to collocation points $(x_i,t_i)$ selected by LHS method in $[-3,3] \times [-5,5]$ of the physics information term $f$.

Then set $N_u = 1000$, $N_{f_u} = 10000$, $d=9$ and $N_{d}=50$, we will give the results of predict $u$ by PINN method and R-PINN method. For the problems, the results use the pre-activation residual connection with $n$=1,2,3 is summarized in Table \ref{compare-F-bugers}.
\begin{table}[htbp]
    \centering
    \caption{Comparison results of PINN and R-PINN methods for the forward problem of variable coefficient Burgers equation}
  \scalebox{0.90}{
    \begin{tabular}{|c|c|c|c|c|}
        \hline
        \multirow{2}{*}{\diagbox{Results}{Methods}} & \multirow{2}{*}{PINN} & \multicolumn{3}{c|}{R-PINN} \\ \cline{3-5}
                                    &      & n=1 & n=2 & n=3 \\ \hline
        error & 5.44$\times 10^{-3}$ & 3.06$\times 10^{-3}$ & 2.07$\times 10^{-3}$ & 3.09$\times 10^{-3}$ \\ \hline
        ERR (\%) & - & 43.82 & 61.92 & 43.26 \\ \hline
    \end{tabular}}
    \label{compare-F-bugers}
\end{table}
The optimization algorithm used in each method here is a combination of 3000 Adam iterations and several L-BFGS iterations. As can be seen from the Table  \ref{compare-F-bugers}, the R-PINN method has a lower error than the PINN method at different parameter $n$ values ($n=1, n=2, n=3$), especially R-PINN($n=2$), which has the lowest error of 2.07$\times 10^{-3}$. Compared with PINN, the R-PINN method significantly reduces the error, especially R-PINN($n=2$) has the highest error reduction rate of 61.92\%.

\begin{figure}[htbp]
\includegraphics[width=0.9\textwidth]{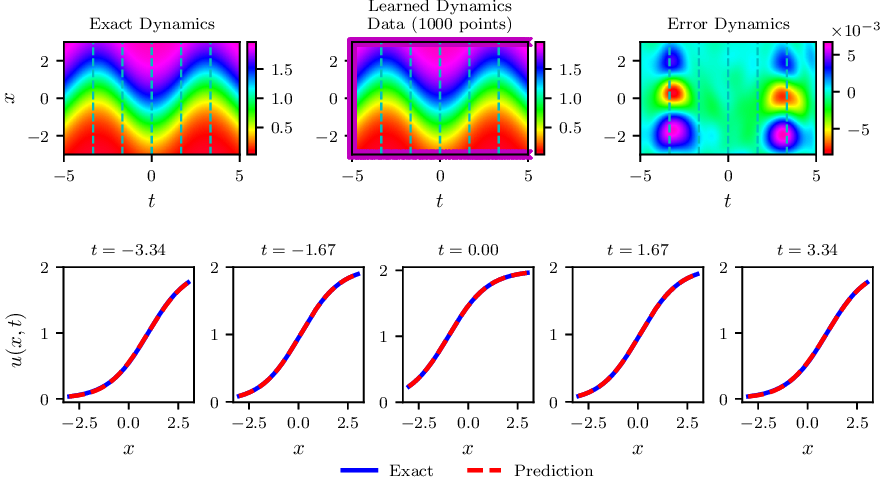}
\centering
\caption{(Color online) The whole drawing is divided into two parts: the upper part of the figure is the dynamic image to compare the exact solutions and predicted solutions of the variable coefficient Burgers equations, and the errors between them; the lower part displays a time slice at a specific point in time.  \textquotedblleft Exact Dynamics" represents the exact values within the solution domain, while \textquotedblleft Learned Dynamics Data" represents the predicted values obtained from R-PINN. Initial and boundary points are indicated by purple \textquotedblleft x" markers. }
\label{F_pred_u_vcbugers}
\end{figure}

\begin{figure}[htbp]
\includegraphics[width=0.6\textwidth]{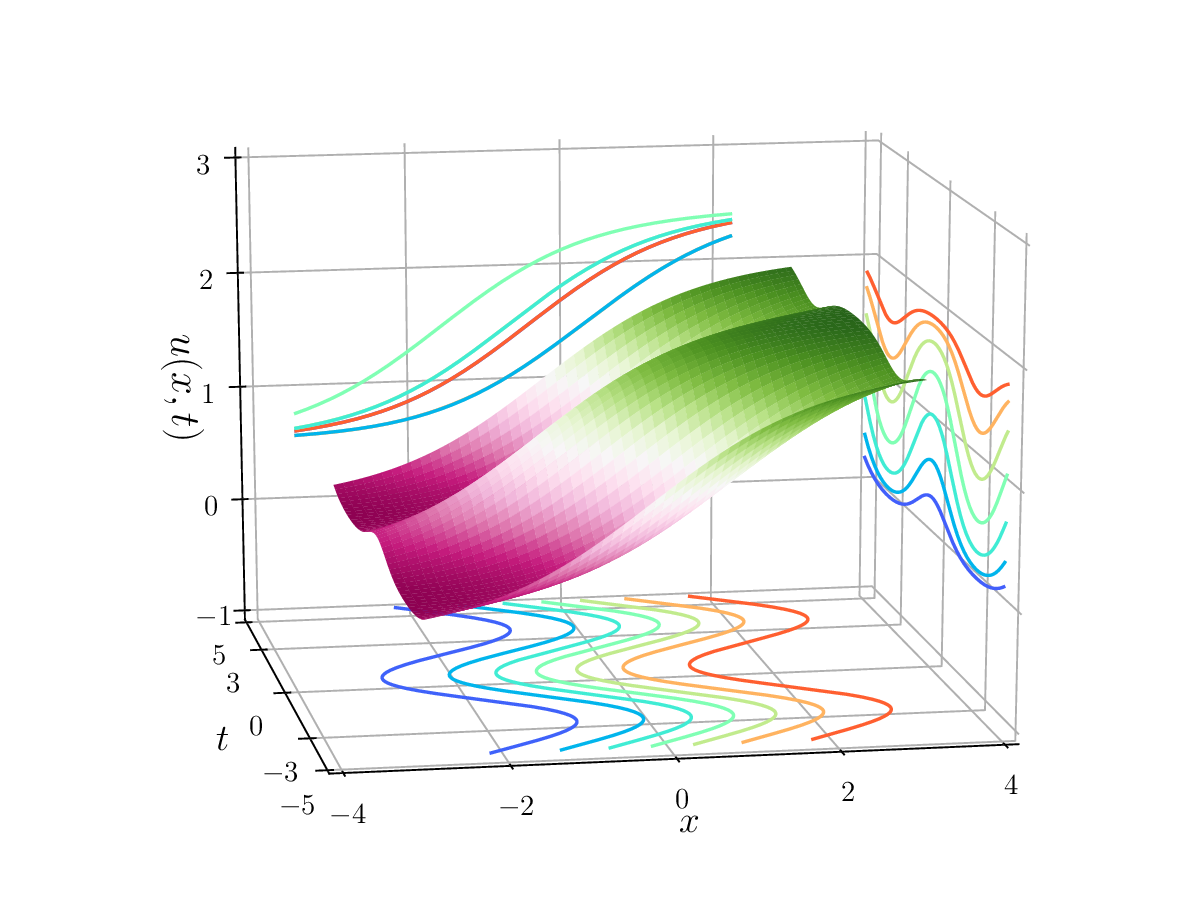}
\centering
\caption{(Color online) 3D surface map and contour projections along the axes of the predicted solution for the variable coefficient Burgers equation via  R-PINN. }
\label{F_pred_u3d_vcbugers}
\end{figure}

Next, we use the R-PINN method with \(n=2\) as an example to provide detailed experimental results. After 3,000 iterations with Adam optimization and 205 iterations with L-BFGS, the error of the predicted solution \(u\) is \(2.07 \times 10^{-3}\). Figure \ref{F_pred_u_vcbugers} shows a comparison between the exact solution and the model's predicted solution, including their dynamic evolution over time and space. The figure also displays the error between the exact and predicted solutions, as well as cross-sectional comparisons at specific time points.
The 3D map of the predicted solution which contains a 3D surface map and contour projections along the axes  is shown in Figure \ref{F_pred_u3d_vcbugers}.


\section{Results-Inverse problems of variable coefficient PDEs}

This section presents the results of numerical experiments using the R-gPINN method to solve variable coefficient inverse problems. Our investigation covers \((1+1)\)-dimensional variable coefficient equations, such as the variable coefficient Burgers equation, the variable coefficient KdV equation, and the variable coefficient Sine-Gordon equation, as well as higher-dimensional variable coefficient KP equations.

To provide a detailed and intuitive display of the numerical results obtained by R-gPINN, we employ a pre-activated residual network with a single hidden layer within each residual block to solve data-driven inverse problems across various equations. For the  \((1+1)\)-dimensional variable coefficient equations problem, once the computational domain \(\Omega \times \left[t_0, t_1\right]\) is established, MATLAB is used to discretize the domain into \(N_x=N_t=1001\) equidistant points in the spatial and temporal directions, respectively. From this discretized domain, we select \(N_{u}=1000\) points corresponding to the initial and boundary conditions, as well as \(N_{u_{in}}=1000\) discrete points within the domain. Additionally, \(N_v=500\) points are chosen along the temporal boundary to serve as the endpoint values for the coefficient \(V(t)\)=\(v(t)\).

In each example, we present the comparison results of the R-PINN, gPINN, and R-gPINN methods for data-driven discovery. Furthermore, we evaluate the impact of different residual connections with 1, 2, and 3 hidden layers on the performance of the R-gPINN method in solving the coefficient discovery problem.

\subsection{Data-driven inverse problems of the variable coefficient Burgers equation}

We employ a 9-layer feedforward neural network with 50 neurons per layer, incorporating pre-activated residual units with a single hidden layer within each residual block, to solve both linear and nonlinear coefficient discovery problems for the variable coefficient Burgers equation.

 \subsubsection{Data-driven discovery of linear coefficient $v(t)=t$ of the variable coefficient Burgers equation}

For \(\beta(t) = t\) and \(k_{1} = m = 1\) in the solution \eqref{eq_3.7}, the variable coefficient Burgers equation has the exact soliton solution:
\begin{equation}\label{vcbuger-ub2}
	u_{b2}(x,t) = 2\frac{e^{x-\frac{1}{2}t^{2}}}{1+e^{x-\frac{1}{2}t^{2}}}.
\end{equation}

We consider the exact coefficient function \(v_{\text{exact}}(x,t) = t\) over the training region \(\Omega = [-2, 2]\) and \(\left[t_0, t_1\right] = [-1, 1]\). Under these conditions, the variable coefficient Burgers equation is given by:
\begin{equation}\label{eq_bugers_t}
\begin{cases}
	u_{t} + v(x,t)uu_{x} + v(x,t)u_{xx} = 0, \\
	u(x,t_0) = u_{b2}(x,-1), \\
	u(x_l,t) = u_{b2}(-2,t), \quad u(x_r,t) = u_{b2}(2,t), \quad (x,t) \in [-2, 2] \times [-1, 1], \\
	u(x_{in}, t_{in}) = u_{b2}(x_{in}, t_{in}), \quad (x_{in}, t_{in}) \in [-2, 2] \times [-1, 1], \\
	v(x,t_{e}) = v_{\text{exact}}(x,t_{e}), \quad t_{e} = t_{0}/t_{1}.
\end{cases}
\end{equation}

\begin{figure}[htbp]
	\centering
	\includegraphics[width=0.8\textwidth]{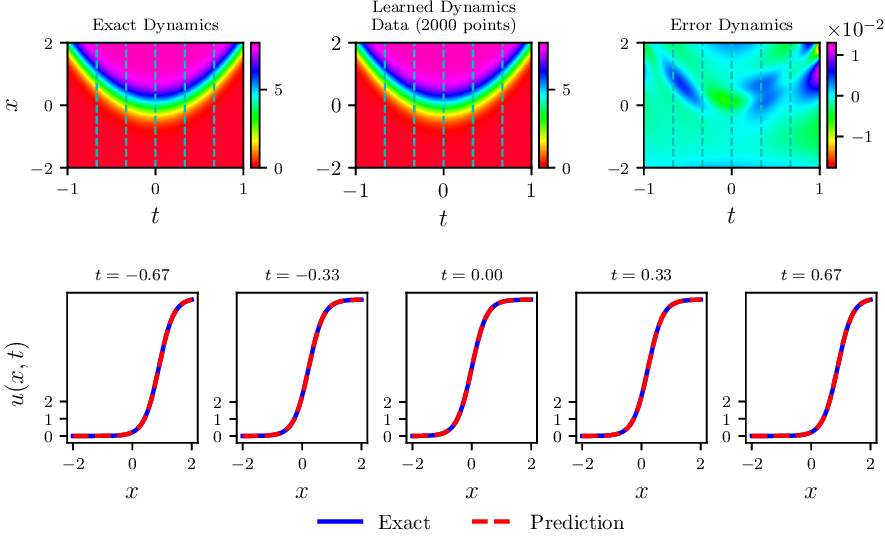}
\caption{(Color online) Data-driven discovery of the linear coefficient \(v(t) = t\) in the variable coefficient Burgers equation: the upper part of the figure shows the dynamic evolution of the exact solutions, predicted solutions, and their corresponding errors; the lower part displays  the time slices at specific points in time.}
	\label{fig_pred_u_buger_t}
\end{figure}

We employ the LHS method to select \(N_f = 10{,}000\) collocation points within the domain \([-2, 2] \times [-1, 1]\). The loss function reaches \(2 \times 10^{-6}\) after $8{,}412$ L-BFGS iterations, with a total training time of 584.4 seconds. We obtain the predict solution \(\hat{u}\) and predict function \(\hat{v}(t)\). The prediction error for \(\hat{u}\) is \(1.08 \times 10^{-4}\), and the error for \(\hat{v}(t)\) is \(6.61 \times 10^{-4}\). The predicted solution \(\hat{u}\) obtained through the R-gPINN method is shown in Figure \ref{fig_pred_u_buger_t}.

\begin{figure}[htbp]\label{fig_U_pred_3Dburgers_cost}
\centering
\subfigure[]{\label{U_pred_3D_bugers_t}
\begin{minipage}[t]{0.45\textwidth}
\centering
\includegraphics[width=\textwidth]{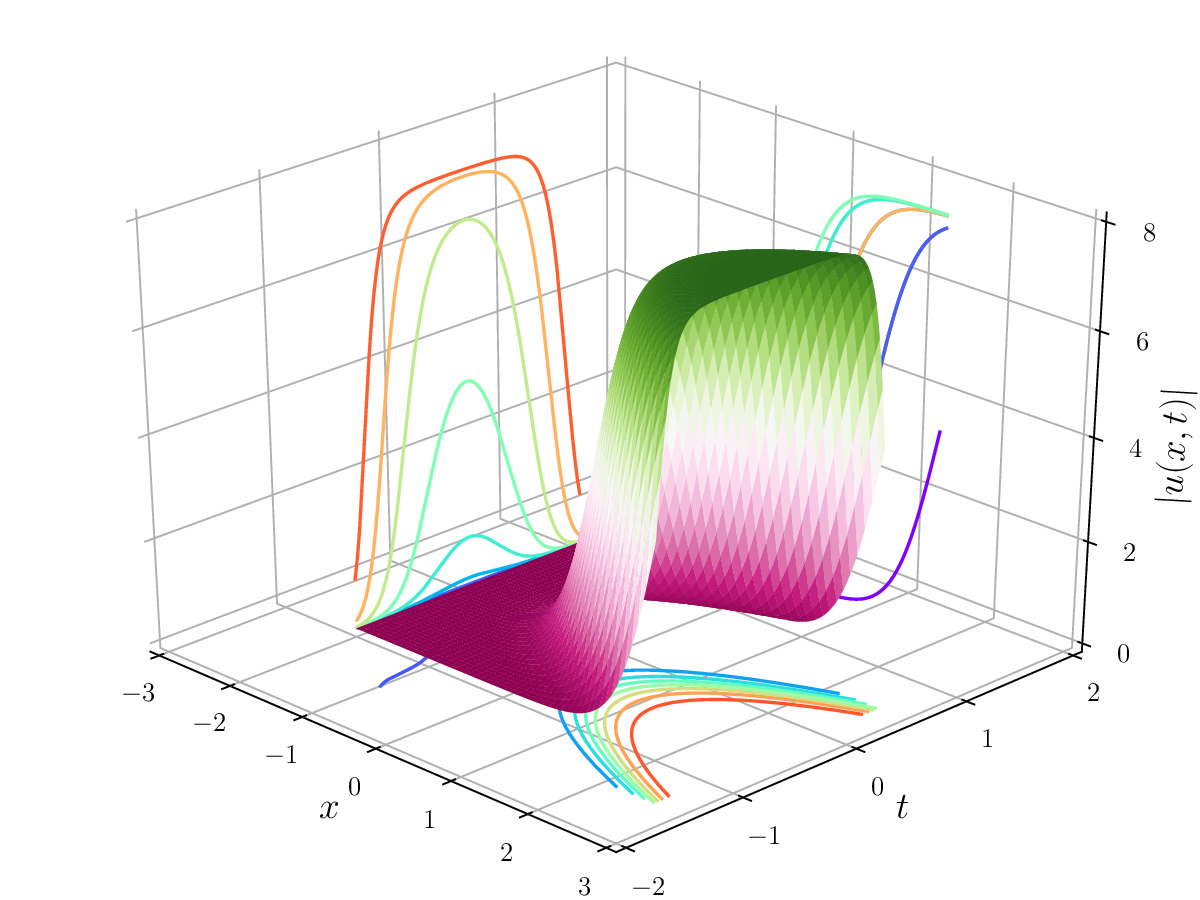}
\end{minipage}
}%
\subfigure[]{\label{fig_comparison_bugers_tv_error1}
\begin{minipage}[t]{0.45\textwidth}
\centering
\includegraphics[width=\textwidth]{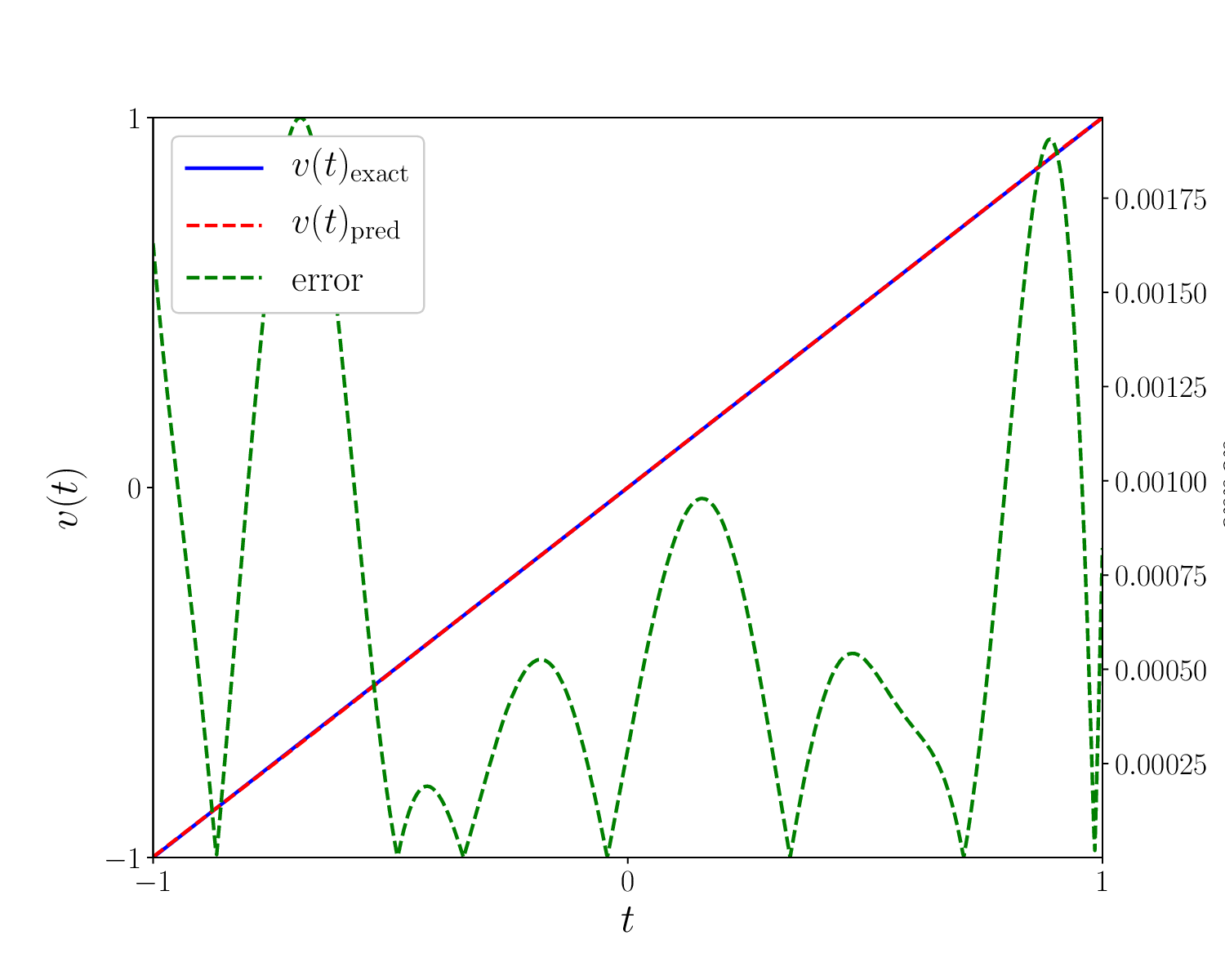}
\end{minipage}%
}%
\caption{ (Color online) (a) 3D plot: The predicted solution \(\hat{u}\) within the region \([-2, 2] \times [-1, 1]\). (b) Data-driven discovery of the variable coefficient \(V(t)\) for the variable coefficient Burgers equation using R-gPINN and comparison of the error between the learned and exact variable coefficient \(V(t)\) over the interval \([-1, 1]\).}
\end{figure}

The 3D plot of the predicted solution \(\hat{u}\) is shown in Figure \ref{U_pred_3D_bugers_t}. By setting \(v(t) = v(x_{i}, t)\), where \(x_{i}\) refers to the \(i\)-th equidistant point in the \(x\) direction, we compare the error \(\text{error} = |v_{\text{exact}} - v_{\text{predict}}|\) between the learned and exact variable coefficients \(v(t)\) over the interval \([-1, 1]\), as illustrated in Figure \ref{fig_comparison_bugers_tv_error1}.

\begin{table}[htbp]
	\centering
\caption{Comparison results of R-PINN, gPINN and R-gPINN methods for the data-driven discovery of the linear coefficient \( v(t) = t \) in the variable coefficient Burgers equation.}
\scalebox{0.90}{
	\begin{tabular}{|c|c|c|c|c|c|}
		\hline
	\multirow{2}{*}{Results} & \multicolumn{3}{c|}{error} & \multirow{2}{*}{$ERR_{13}$(\%)} & \multirow{2}{*}{$ERR_{23}$ (\%)} \\\cline{2-4}
                           & R-PINN &  gPINN &  R-gPINN &  &   \\                
		\hline
		\( u \) & 2.49$\times 10^{-4}$ & \(2.00 \times 10^{-4}\)& \(1.08 \times 10^{-4}\) & 56.39 & 45.66 \\
		\hline
		\( v \)  &1.60$\times 10^{-2}$   & \(8.47 \times 10^{-4}\)& \(6.61 \times 10^{-4}\)& 95.86 & 21.95  \\
		\hline
	\end{tabular}}
	\label{compare-bugers-t}
\end{table}
Table \ref{compare-bugers-t} presents a comparison of the R-PINN, gPINN, and R-gPINN methods for the data-driven discovery of the linear coefficient \( v(t) = t \) in the variable coefficient Burgers equation. The results show that the R-gPINN method achieved the lowest error$_{u}$ and error$_{v}$ values, with \(1.08 \times 10^{-4}\) and \(6.61 \times 10^{-4}\) respectively, outperforming both the R-PINN and gPINN methods. Specifically, the R-gPINN method reduced the error$_{u}$ by 56.39\% and 45.66\% compared to the R-PINN and gPINN methods, respectively. For error$_{v}$, the R-gPINN method achieved a reduction of 95.86\% compared to the R-PINN method and 21.95\% compared to the gPINN method. These results indicate that the R-gPINN method provides significantly better accuracy in solving this problem compared to the other two methods.

Here, two types of residual connections were designed: pre-activation and post-activation connections. To evaluate the impact of these different connection methods on the results, we tested the outcomes using residual blocks with 1, 2, and 3 hidden layers for each type of residual connection. For example, pre3 means a pre-activated residual network with a single hidden layer in each residual block. post3 means a post-activated residual network with three hidden layer in each residual block. The performance is evaluated using two key error metrics, error$_{u}$ and error$_{v}$, which quantify the accuracy of the predicted solutions and predicted function. 
\begin{table}[h!]
\centering
\caption{Evaluation of the impact of different residual  connection with 1, 2, and 3 hidden layers on the performance of the R-gPINN method in solving the linear coefficient  \( v(t) = t \)  discovery  problem for the variable coefficient Burgers equation.}
\scalebox{0.90}{
\begin{tabular}{|c|c|c|c|c|c|c|}
\hline
\multirow{2}{*}{\diagbox{Results}{Methods}} &  \multicolumn{6}{c|}{R-gPINN} \\ \cline{2-7}
                                            & pre1 & pre2 & pre3&post1&post2&post3 \\ \hline
 error$_{u}$  & 1.08$\times 10^{-4}$    & 1.30$\times 10^{-4}$ & 6.82$\times 10^{-5}$  & 1.02$\times 10^{-4}$ & 9.68$\times 10^{-5}$ & 8.19$\times 10^{-5}$ \\ \hline
 error$_{v}$ & 6.61$\times 10^{-4}$  & 7.63$\times 10^{-4}$   & 3.38$\times 10^{-4}$ & 6.80$\times 10^{-4}$ & 1.01$\times 10^{-3}$ & 8.04$\times 10^{-4}$ \\ \hline
\end{tabular}}
\label{sum-bugers-t}
\end{table}
Table \ref{sum-bugers-t} evaluates the effect of different residual connection types and the number of hidden layers on the performance of the R-gPINN method in solving the linear coefficient \( v(t) = t \) discovery problem for the variable coefficient Burgers equation. The results indicate that, for error$_{u}$, the best performance is achieved with the R-gPINN(pre3) configuration, which has three hidden layers and a pre-activation residual connection, resulting in the lowest error of \(6.82 \times 10^{-5}\). For error$_{v}$, the lowest value is also achieved with the R-gPINN(pre3) method, with an error of \(3.38 \times 10^{-4}\). Overall, the pre-activation residual connections tend to outperform the post-activation connections in this context, and increasing the number of hidden layers generally improves the model's performance, with three hidden layers providing the best results for both error metrics.


\subsubsection{Data-driven discovery of nonlinear coefficient $v(t)=cos(t)$ of the variable coefficient Burgers equation}

When \(\beta(t) = \cos(t)\) and \(k_{1} = m = 1\) are applied in solution \eqref{eq_3.7}, the variable coefficient Burgers equation yields the exact soliton solution:
\begin{equation}\label{eq_buger_ub3}
	u_{b3}(x,t) = 2 \frac{e^{x - \sin(t)}}{1 + e^{x - \sin(t)}}.
\end{equation}

\begin{figure}[htbp]
	\centering
		\includegraphics[width=0.8\textwidth]{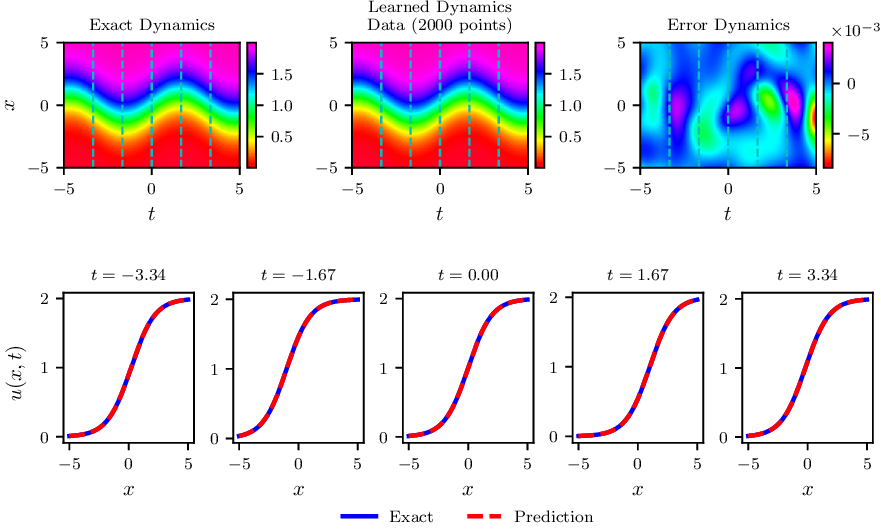}
		\caption{(Color online) Data-driven discovery of the nonlinear coefficient \(v(t) = \cos(t)\) in the variable coefficient Burgers equation: the upper part of the figure shows the dynamic evolution of the exact solutions, predicted solutions, and their corresponding errors; the lower part displays  the time slices at specific points in time.}
\label{fig_pred_u_buger_cost}
\end{figure}

The goal function is \(v_{exact}(x,t) = \cos(t)\), with the training region defined as \(\Omega \times [t_0, t_1] = [-5,5] \times [-5,5]\). The variable coefficient Burgers equation with these conditions is given by:
\begin{equation}
\begin{cases}
		u_{t} + v(x,t) u u_{x} + v(x,t) u_{xx} = 0,  \\
		u(x,t_0) = u_{b3}(x,-5), \\
		u(x_l,t) = u_{b3}(-5,t), \ u(x_r,t)= u_{b3}(5,t), \ (x,t) \in [-5,5] \times [-5,5],\\
		u(x_{in},t_{in}) = u_{b3}(x_{in},t_{in}), \ (x_{in},t_{in}) \in [-5,5] \times [-5,5],\\
		v(x,t_{e}) = v_{exact}(x,t_{e}), \ t_{e} = t_{0}/t_{1}.
	\label{eq_bugers_cost}
	\end{cases}
\end{equation}
Here, \((x_{in},t_{in})\) are grid points within the region \([-5,5] \times [-5,5]\). Using the LHS method, 10,000 collocation points were selected within the domain. Then the error of predicted solution \(\hat{u}\) is \(2.48 \times 10^{-4}\)  underwent 3,000 Adam iterations followed by 2,294 L-BFGS optimization steps, totaling a training time of 529.36 seconds. The  dynamic evolution of the exact solutions, predicted solutions,  errors and their time slices are depicted in Figure \ref{fig_pred_u_buger_cost}.

\begin{figure}[htbp]\label{fig_U_pred_3Dburgers_cost}
\subfigure[]{\label{U_pred_3Dburgers_cost}
\begin{minipage}[t]{0.45\textwidth}
\centering
\includegraphics[width=\textwidth]{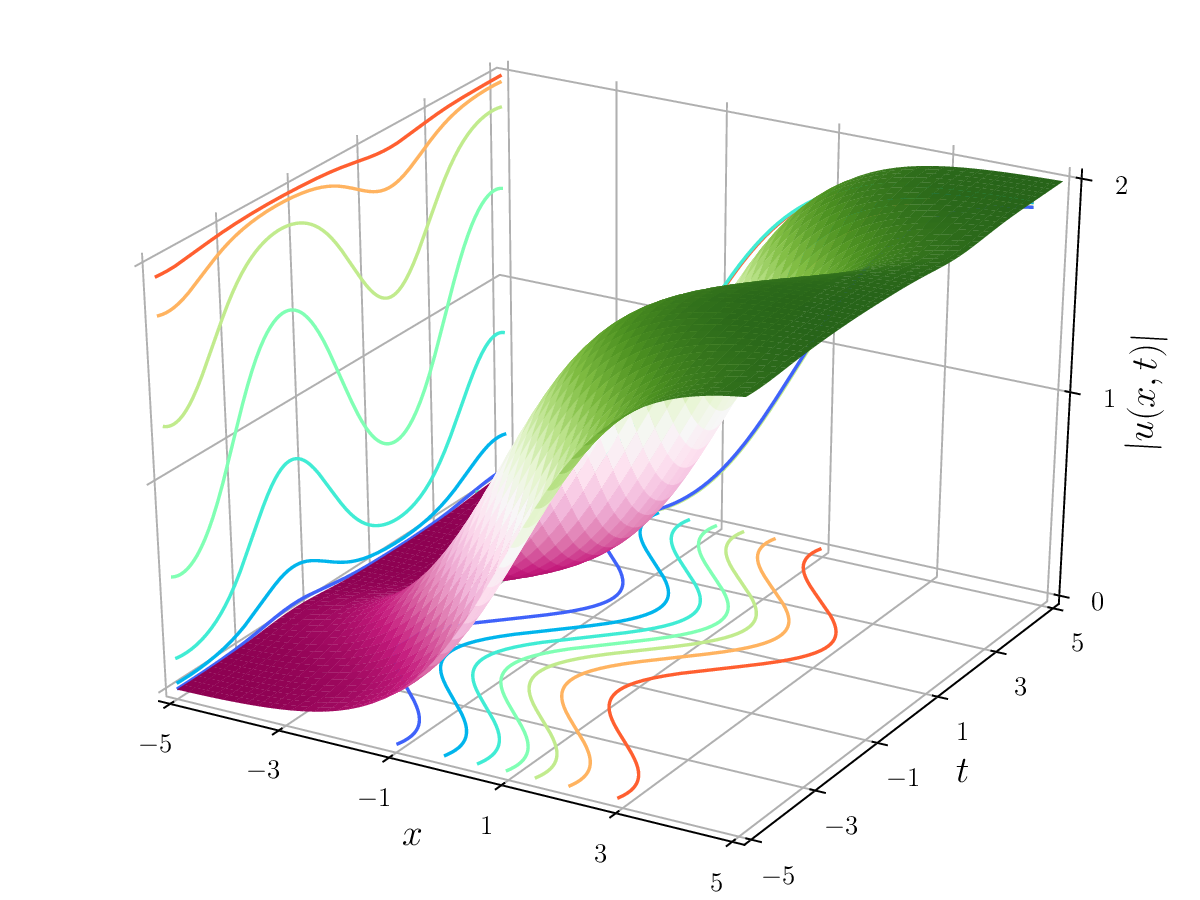}
\end{minipage}
}%
\subfigure[]{\label{fig_comparison_bugers_costv}
\begin{minipage}[t]{0.45\textwidth}
\centering
\includegraphics[width=\textwidth]{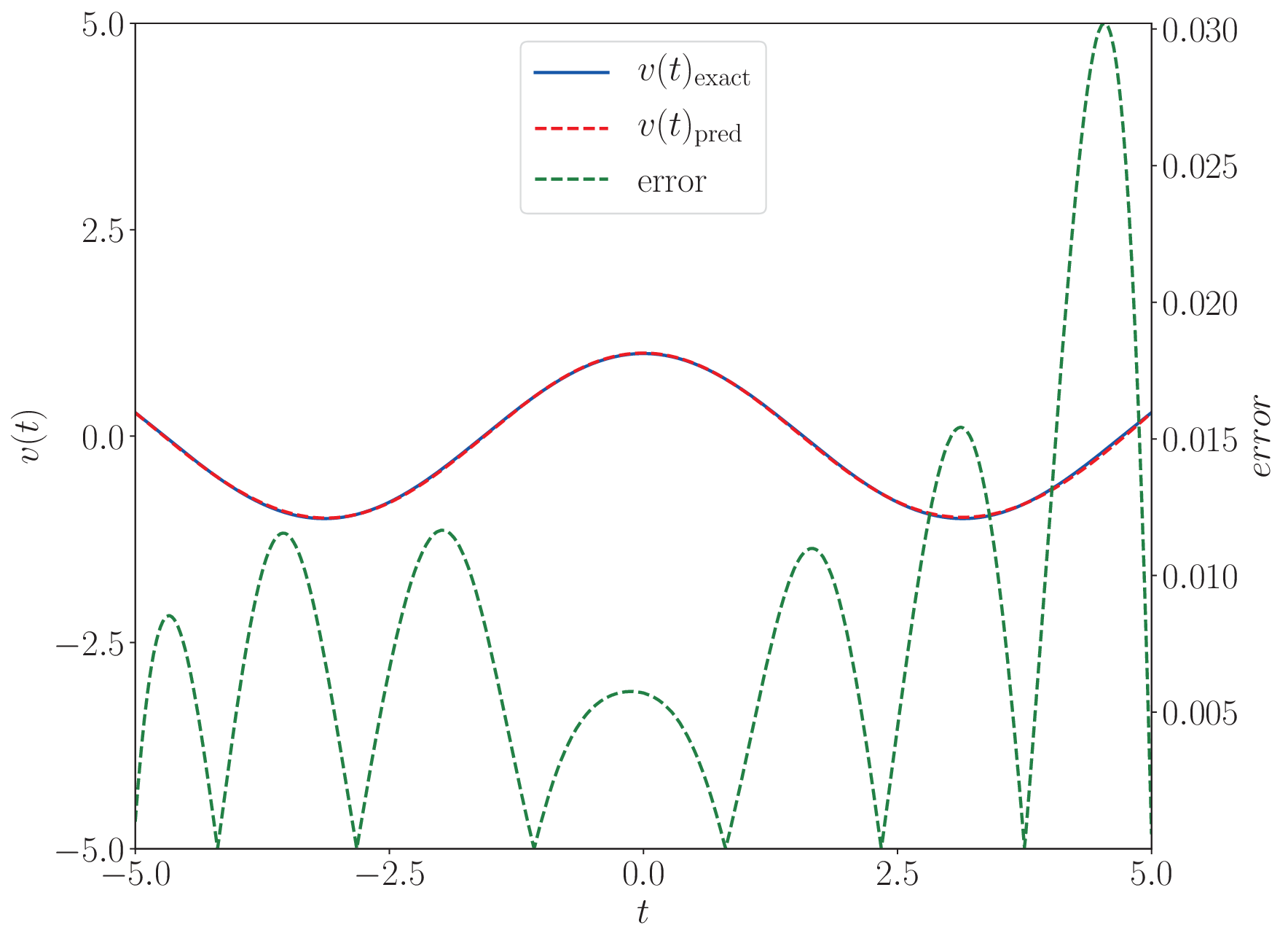}
\end{minipage}%
}%
\centering
	\caption{ (Color online) (a) 3D plot: The predicted solution \(\hat{u}\) within the region \([-5,5] \times [-5,5]\). (b) Data-driven discovery of the variable coefficient \(V(t)\) for the variable coefficient Burgers equation using R-gPINN and comparison of the error between the learned and exact variable coefficient \(V(t)\) over the interval \([-5, 5]\).}
\end{figure}

The 3D plot of the predicted solution \(\hat{u}\) within the region \([-5,5] \times [-5,5]\) is shown in \ref{U_pred_3Dburgers_cost}. The comparison of the error between the learned and exact variable coefficient \(v(t)\) over the interval \([-1,1]\) is shown in Figure \ref{fig_comparison_bugers_costv}, with an error for \(\hat{v}(t)\) of \(3.79 \times 10^{-3}\).

For a direct comparison of prediction accuracy, the accuracy comparison of R-PINN, gPINN and R-gPINN methods for the data-driven discovery of the nonlinear coefficient \( v(t) = \cos(t) \) in the variable coefficient Burgers equation is presented in Table \ref{compare-bugers-cost}.
\begin{table}[htbp]
	\centering
\caption{Comparison results of R-PINN, gPINN and R-gPINN methods for the data-driven discovery of the nonlinear coefficient \( v(t) = \cos(t) \) in the variable coefficient Burgers equation}
  \scalebox{0.9}{
	\begin{tabular}{|c|c|c|c|c|c|}
		\hline
		\multirow{2}{*}{Results} & \multicolumn{3}{c|}{error} & \multirow{2}{*}{$ERR_{13}$(\%)} & \multirow{2}{*}{$ERR_{23}$ (\%)} \\\cline{2-4}
                           & R-PINN &  gPINN &  R-gPINN &  &   \\                
		\hline
		\( u \) & 5.89$\times 10^{-4}$ & \(1.00\times 10^{-3}\) & \(2.48 \times 10^{-4}\)& 57.84 & 75.25  \\
		\hline
		\( v \) &4.48$\times 10^{-2}$ & \(1.73\times 10^{-2}\) & \(3.79 \times 10^{-3}\) & 91.53 & 78.12 \\
		\hline
	\end{tabular}}
	\label{compare-bugers-cost}
\end{table}
Table \ref{compare-bugers-cost} compares the results of the R-PINN, gPINN and R-gPINN methods for the inverse problem of the variable coefficient Burgers equation, where the goal function is \(v(t) = \cos(t)\). The R-gPINN method shows a significant improvement over the another method. For example, the error in predicting \(u\) is reduced by 75.25\%, with R-gPINN achieving an error of \(2.48 \times 10^{-4}\) compared to \(1.00 \times 10^{-3}\) for gPINN. Similarly, the error in predicting \(v\) is reduced by 78.12\%, with R-gPINN achieving an error of \(3.79 \times 10^{-3}\) compared to \(1.73 \times 10^{-2}\) for gPINN. Overall, the R-gPINN method demonstrates superior accuracy in solving the inverse problem of the variable coefficient Burgers equation.

\begin{table}[h!]
\centering
\caption{Evaluation of the impact of  different residual  connection with 1, 2, and 3 hidden layers on the performance of the R-gPINN method in solving the  nonlinear coefficient \( v(t) = \cos(t) \) discovery  problem for the variable coefficient Burgers equation.}
  \scalebox{0.9}{
\begin{tabular}{|c|c|c|c|c|c|c|c|}
\hline
\multirow{2}{*}{\diagbox{Results}{Methods}} &  \multicolumn{6}{c|}{R-gPINN} \\ \cline{2-7}
                                            & pre1 & pre2 & pre3&post1&post2&post3 \\ \hline
error$_{u}$ & 2.48$\times 10^{-4}$   & 4.77$\times 10^{-4}$    & 9.62$\times 10^{-4}$  & 2.41$\times 10^{-4}$            & 5.88$\times 10^{-4}$        & 6.32$\times 10^{-4}$  \\ \hline
error$_{v}$ & 3.79$\times 10^{-3}$   & 8.37$\times 10^{-3}$     & 1.82$\times 10^{-2}$  & 4.01$\times 10^{-3}$           & 1.13$\times 10^{-2}$       & 1.22$\times 10^{-2}$  \\ \hline
\end{tabular}}
\label{sum-bugers-cost}
\end{table}
Table \ref{sum-bugers-cost} evaluates the impact of different residual connections with 1, 2, and 3 hidden layers on the performance of the R-gPINN method in discovering the nonlinear coefficient \( v(t) = \cos(t) \) for the variable coefficient Burgers equation. For error$_{u}$, the table shows that the R-gPINN method with pre-activation residual connections performs better than with post-activation connections. Specifically, the configuration with 1 hidden layer (pre1) achieves the lowest error of \(2.48 \times 10^{-4}\). Errors increase with the number of hidden layers, reaching \(9.62 \times 10^{-4}\) for 3 hidden layers (pre3). Similarly, for post-activation connections, the method with 1 hidden layer (post1) shows the lowest error of \(2.41 \times 10^{-4}\), but the error rises to \(6.32 \times 10^{-4}\) with 3 hidden layers (post3).
Regarding error$_{v}$, the trend is similar. The pre-activation configuration with 1 hidden layer (pre1) yields the lowest error of \(3.79 \times 10^{-3}\). This error increases with more hidden layers, reaching \(1.82 \times 10^{-2}\) for 3 hidden layers (pre3). For post-activation connections, the method with 1 hidden layer (post1) also performs best with an error of \(4.01 \times 10^{-3}\), which increases to \(1.22 \times 10^{-2}\) with 3 hidden layers (post3). In summary, both error$_{u}$ and error$_{v}$ are minimized when using pre-activation residual connections with fewer hidden layers, indicating that this configuration is more effective for solving the nonlinear coefficient discovery problem in the variable coefficient Burgers equation.

\subsection{Data-driven inverse problems of the variable coefficient KdV equation}

The variable coefficient KdV equation models various physical phenomena, including solitary waves in varying water depths, internal gravity waves in lakes, ion acoustic waves in plasmas, and wave damping. Grimshaw derived a simplified form of the variable coefficient KdV equation to describe slowly varying solitary waves \cite{Grimshaw-1979}, which is given by:
\begin{equation}\label{vc-kdv}
u_t + v(t)uu_x + g(t)u_{xxx} = 0.
\end{equation}

By setting \(g(t) = cv(t)\), where \(c\) is an arbitrary constant, the equation \eqref{vc-kdv} admits an exact solution provided by \cite{fan-2002}:
\begin{equation}\label{vc-kdv-s}
u_{k}(x, t) = 3 c \alpha^2 \operatorname{sech}^2 \left[\frac{1}{2} \alpha \left(x - c \alpha^2 \int f(t) \, dt \right) \right],
\end{equation}
where \(\alpha\) is a free parameter. Setting \( c = \alpha = 1 \) simplifies the solution \eqref{vc-kdv-s} to depend solely on the variable coefficient \( v(t) \).
We will investigate the variable coefficient \( v(t) \) of the variable coefficient KdV equation for both linear (\( v(t) = t \)) and nonlinear (\( v(t) = \cos(t) \)) cases. The training domain is chosen as \([-5,5] \times [-5,5]\), with 20,000 sampling points selected within this domain using the LHS method. For the experiments, the parameter \( d \) is set to 9, and \( N_d \) is set to 40. The results will include predictions for \( u \) and \( v \) using the R-gPINN method, along with a comparison of the relative \(\mathbb{L}_2\) norm errors between the R-gPINN, gPINN, and R-gPINN methods.

 \subsubsection{Data-driven discovery of linear coefficient $v(t)=t$ of the variable coefficient KdV equation}
 
When the coefficient \(v(t)\) is chosen as a linear function, such as \(v(t) = t\), the exact solution of the variable coefficient KdV equation \eqref{vc-kdv} becomes:
\begin{equation}\label{vc-kdv-st}
u_{k1} = 3 \operatorname{sech}^2 \left[\frac{1}{4} \left(t^2 - 2x\right) \right].
\end{equation}

The goal function is \(v_{exact}(x,t) = t\), with the training region defined as \(\Omega \times [t_0, t_1] = [-5,5] \times [-5,5]\). The variable coefficient KdV equation under these conditions is given by:
\begin{equation}\label{i-vckdv}
\begin{cases}
    u_t + v(t) uu_x + v(t) u_{xxx} = 0,  \\
    u(x, t_0) = u_{k1}(x, -5), \\
    u(x_l, t) = u_{k1}(-5, t), \quad u(x_r, t) = u_{k1}(5, t), \quad (x, t) \in [-5,5] \times [-5,5], \\
    u(x_{in}, t_{in}) = u_{k1}(x_{in}, t_{in}), \quad (x_{in}, t_{in}) \in [-5,5] \times [-5,5], \\
    v(x, t_{e}) = v_{exact}(x, t_{e}), \quad t_{e} = t_{0}/t_{1}.
\end{cases}
\end{equation}

\begin{figure}[htbp]
	\centering
		\includegraphics[width=0.8\textwidth]{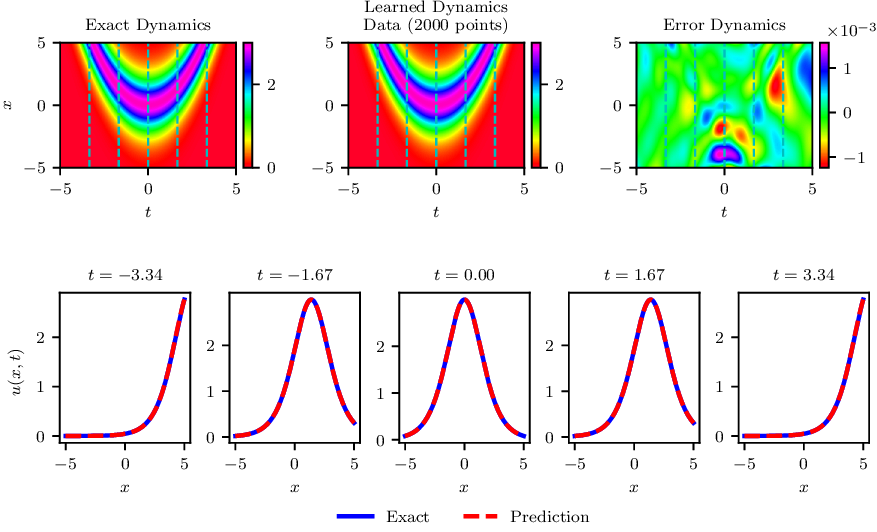}
		\caption{(Color online) Data-driven discovery of the linear coefficient \(v(t) = t\) in the variable coefficient KdV equation: the upper part of the figure shows the dynamic evolution of the exact solutions, predicted solutions, and their corresponding errors; the lower part displays  the time slices at specific points in time.}
		\label{fig_pred_u_kdv_t}
\end{figure}

 The objective loss function value descend to 1$\times 10^{-6}$ via 3727 L-BFGS iterations, with a total training time of 528.03 seconds. The predicted solution \(\hat{u}\) obtained via the R-gPINN method is depicted in Figure \ref{fig_pred_u_kdv_t}, with an error of \(3.13 \times 10^{-4}\). The figure below shows a comparison between the "exact solution" and the "predicted solution" of the variable coefficient KdV equation, as well as their errors.
By comparing the prediction solution of the R-gPINN method with the exact solution of the equation, the accuracy of the model can be evaluated more intuitively, and their performance and error at different time points can be demonstrated through heat maps and slice maps.

\begin{figure}[htbp]
\subfigure[]{\label{fig_3Dut-vckdv}
\begin{minipage}[t]{0.45\textwidth}
\centering
\includegraphics[width=\textwidth]{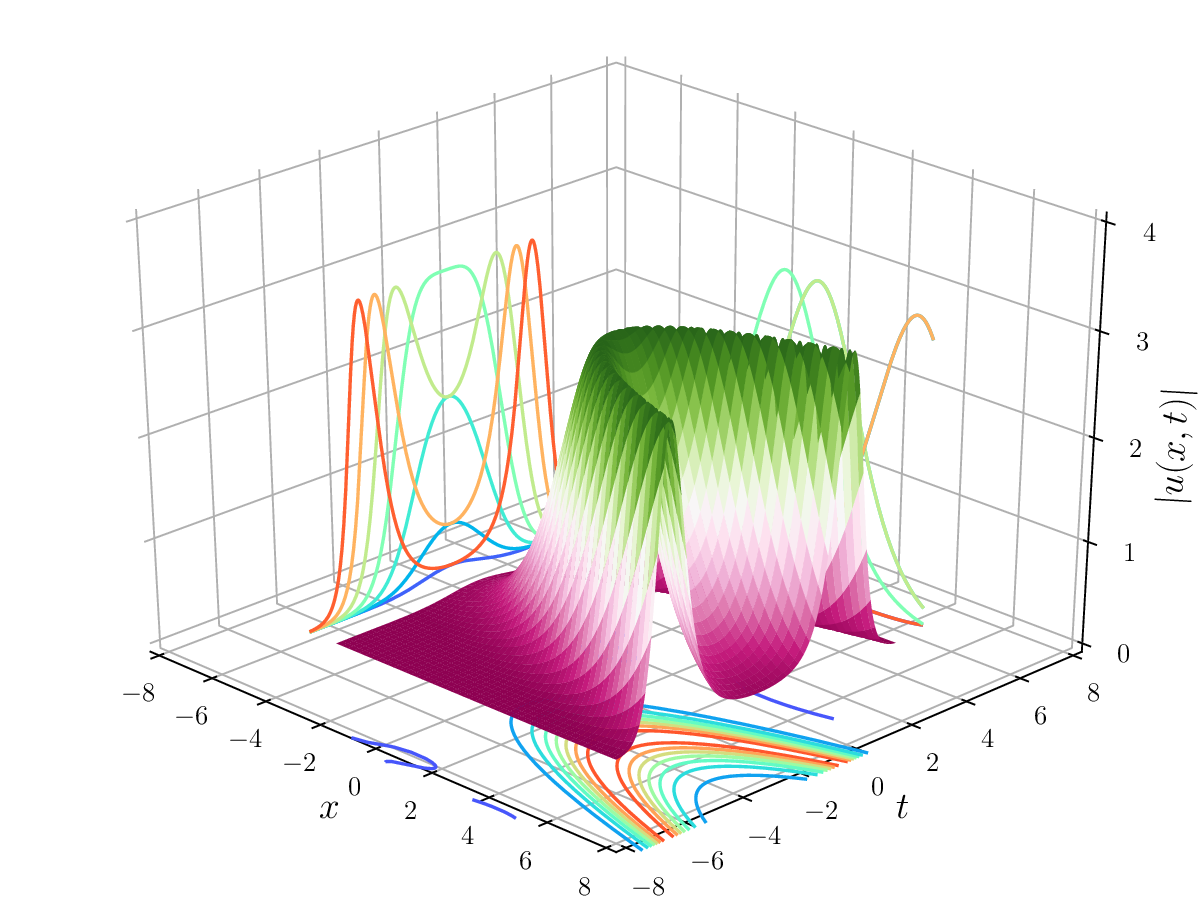}
\end{minipage}
}%
\subfigure[]{\label{fig_comparison_kdv_tv_error1}
\begin{minipage}[t]{0.45\textwidth}
\centering
\includegraphics[width=\textwidth]{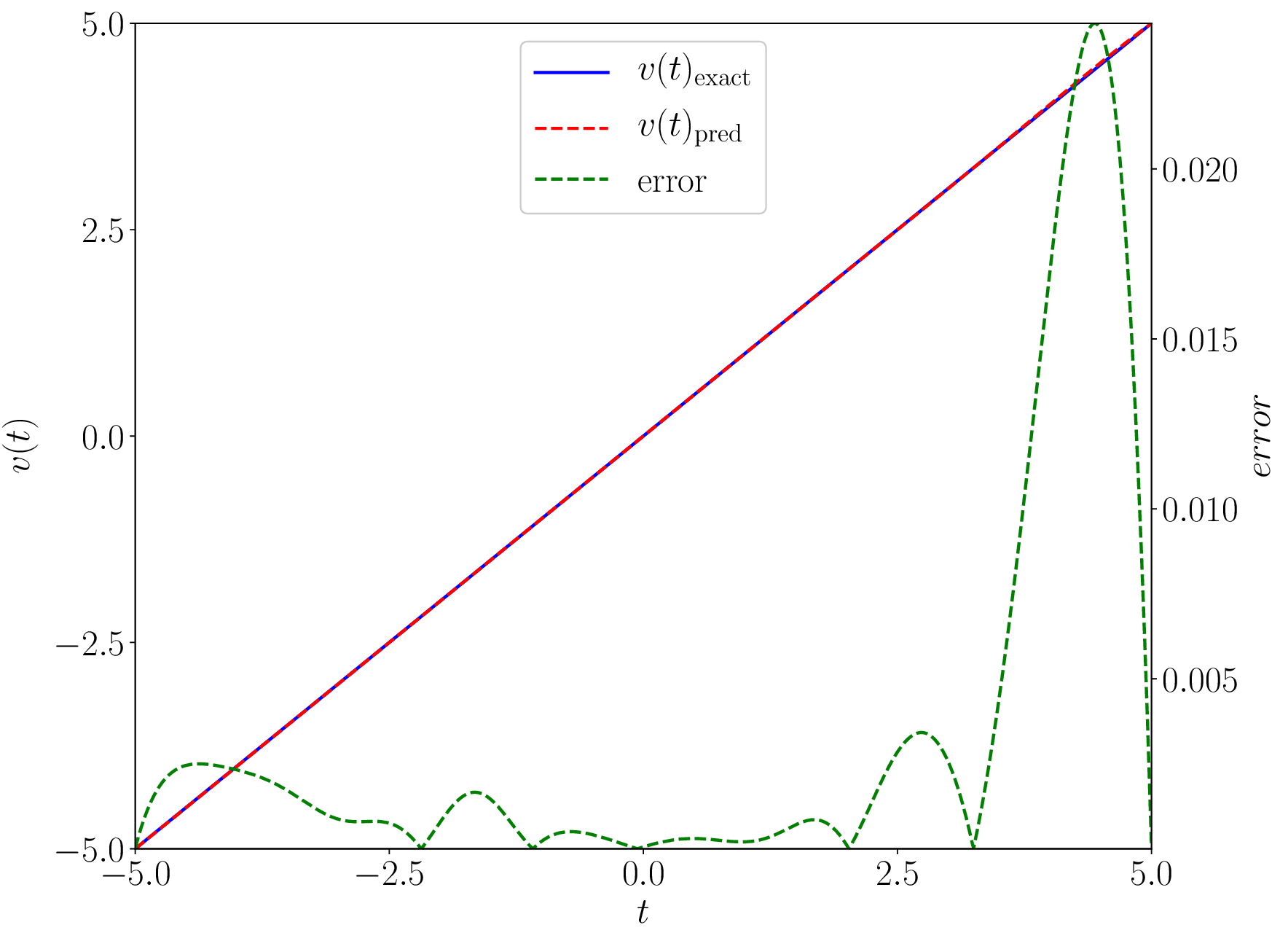}
\end{minipage}%
}%
\centering
\caption{ (Color online) Goal function \(v(x,t) = t\): (a) 3D plot of the predicted solution \(\hat{u}\) of the variable coefficient KdV equation within the region \([-5,5] \times [-5,5]\); (b) Comparison of the error between the learned and exact variable coefficient \(v(t)\) over the interval \([-5,5]\) using the R-gPINN method.}
\end{figure}

The 3D plot of the predicted solution \(\hat{u}\) within the region \([-5,5] \times [-5,5]\) is shown in Figure\ref{fig_3Dut-vckdv}. The comparison of the error between the learned and exact variable coefficient \(v(t)\) over the interval \([-1,1]\) is shown in Figure \ref{fig_comparison_kdv_tv_error1}, with an error for \(\hat{v}(t)\) of \(2.41\times 10^{-3}\).

The Table \ref{compare-kdv-t} compares the results of solving the inverse problem of the variable coefficient KdV equation using the  R-PINN, gPINN and R-gPINN methods, with the target function \(v(t) = t\).
\begin{table}[htbp]
	\centering
	\caption{Comparison results of R-PINN, gPINN and R-gPINN methods for the data-driven discovery of the linear coefficient \( v(t) = t \) in the variable coefficient KdV equation}
  \scalebox{0.9}{
	\begin{tabular}{|c|c|c|c|c|c|}
		\hline
		\multirow{2}{*}{Results} & \multicolumn{3}{c|}{error} & \multirow{2}{*}{$ERR_{13}$(\%)} & \multirow{2}{*}{$ERR_{23}$ (\%)} \\\cline{2-4}
                           & R-PINN &  gPINN &  R-gPINN &  &   \\                
		\hline
	\( u \) &4.56$\times 10^{-4}$  & 3.93$\times 10^{-4}$  & 3.13$\times 10^{-4}$  & 31.43&20.34 \\
		\hline
		\( v \) & 1.55$\times 10^{-2}$  & 2.82$\times 10^{-3}$  & 2.41$\times 10^{-3}$  & 84.43 &14.52  \\
		\hline
	\end{tabular}}
	\label{compare-kdv-t}
\end{table}
 The results in Table \ref{compare-kdv-t} indicate that for the error in \(u\), the gPINN method has an error of \(3.93 \times 10^{-4}\), while the R-gPINN method achieves a smaller error of \(3.13 \times 10^{-4}\), representing a reduction of approximately 20.34\%. For the error in \(v\), the gPINN method has an error of \(2.82 \times 10^{-3}\), whereas the R-gPINN method reduces this to \(2.41 \times 10^{-3}\), a decrease of about 14.52\%. This demonstrates that the R-gPINN method performs better in addressing this inverse problem.

\begin{table}[h!]
\centering
\caption{Evaluation of the impact of  different residual  connection with 1, 2, and 3 hidden layers on the performance of the R-gPINN method in solving the  linear coefficient  \( v(t) = t \)  discovery  problem for the variable coefficient KdV equation.}
  \scalebox{0.9}{
\begin{tabular}{|c|c|c|c|c|c|c|c|}
\hline
\multirow{2}{*}{\diagbox{Results}{Methods}} &  \multicolumn{6}{c|}{R-gPINN} \\ \cline{2-7}
                                            & pre1 & pre2 & pre3&post1&post2&post3 \\ \hline
error$_{u}$ & 3.13$\times 10^{-4}$      & 2.67$\times 10^{-4}$     & 2.99$\times 10^{-4}$   & 3.03$\times 10^{-4}$    & 3.57$\times 10^{-4}$    & 2.47$\times 10^{-4}$   \\ \hline
 error$_{v}$ & 2.41$\times 10^{-3}$       & 2.62$\times 10^{-3}$     & 2.61$\times 10^{-3}$    & 5.78$\times 10^{-3}$   & 3.58$\times 10^{-3}$   & 2.15$\times 10^{-3}$ \\ \hline
\end{tabular}}
\label{sum-kdv-t}
\end{table}
Table \ref{sum-kdv-t} assesses the impact of different residual connections with 1, 2, and 3 hidden layers on the performance of the R-gPINN method for solving the linear coefficient \( v(t) = t \) discovery problem in the variable coefficient KdV equation. For error$_{u}$, the data indicate that the best performance is achieved with the post-activation residual connection and 3 hidden layers (post3), which has the lowest error of \(2.47 \times 10^{-4}\). Among the pre-activation connections, the error$_{u}$ is lowest with 2 hidden layers (pre2), yielding \(2.67 \times 10^{-4}\), but increases with 1 and 3 hidden layers. For post-activation connections, the error$_{u}$ increases with more hidden layers, reaching \(3.57 \times 10^{-4}\) for 2 hidden layers (post2).
Regarding error$_{v}$, the best results are also seen with the post-activation connection and 3 hidden layers (post3), achieving the lowest error of \(2.15 \times 10^{-3}\). The pre-activation connections show the lowest error with 1 hidden layer (pre1) at \(2.41 \times 10^{-3}\), but the error increases with additional hidden layers.
Overall, the R-gPINN method performs optimally for both error$_{u}$ and error$_{v}$ with post-activation connections and 3 hidden layers, suggesting that this configuration provides the most accurate results for solving the linear coefficient discovery problem in the KdV equation.

\subsubsection{Data-driven discovery of nonlinear coefficient $v(t)=\cos(t)$ of the variable coefficient KdV equation}

When the coefficient function is a cosine function, that is, $v(t)=\cos (t)$, the exact solution of the corresponding variable coefficient KdV equation is
$$
u_{k2}=3 \operatorname{sech}^2\left[\frac{1}{2}(x-\sin (t))\right] .
$$
The goal function is \(v_{k2_{exact}}(x,t) = \cos(t)\), with the training region defined as \(\Omega \times [t_0, t_1] = [-5,5] \times [-5,5]\). The variable coefficient Burgers equation with these conditions is given by:
\begin{equation}\label{i-vckdv2}
\begin{cases}
		u_t+v(t)uu_x+v(t)u_{x x x}=0,  \\
		u(x,t_0) = u_{k2}(x,-5), \\
		u(x_l,t) = u_{k2}(-5,t), u(x_r,t)= u_{k2}(5,t),  (x,t)  \in [-5,5] \times [-5,5],\\
		u(x_{in},t_{in})=u_{k2}(x_{in},t_{in}), (x_{in},t_{in}) \in [-5,5] \times [-5,5],\\
		v(x,t_{e})=v_{exact}(x,t_{e}),t_{e}=t_{0}/t_{1}.
	\end{cases}
\end{equation}

\begin{figure}[htbp]
	\centering
		\includegraphics[width=0.8\textwidth]{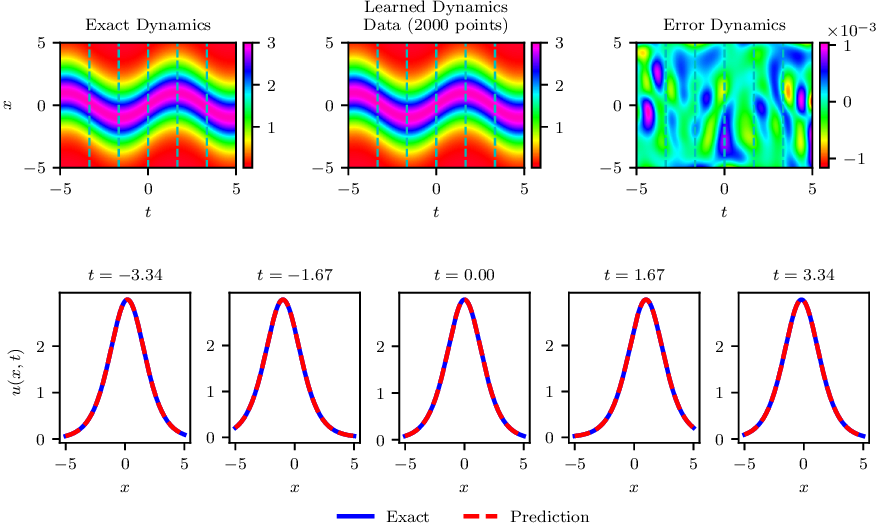}
		\caption{(Color online) Data-driven discovery of the nonlinear coefficient \(v(t) =\cos(t)\) in the variable coefficient KdV equation: the upper part of the figure shows the dynamic evolution of the exact solutions, predicted solutions, and their corresponding errors; the lower part displays  the time slices at specific points in time.}
		\label{fig_pred_u_kdv_cost}
\end{figure}

The predicted solution \(\hat{u}\), obtained using the R-gPINN method after 3000 iterations of Adam optimization and 3808 iterations of L-BFGS optimization, is illustrated in Figure \ref{fig_pred_u_kdv_cost}. The prediction error is \(2.68 \times 10^{-4}\). The figure compares the exact solution with the predicted solution for the variable coefficient KdV equation, along with the corresponding errors.
By comparing the prediction solution of the R-gPINN method with the exact solution of the equation, the accuracy of the model can be evaluated more intuitively, and their performance and error at different time points can be demonstrated through heat maps and slice maps.

\begin{figure}[htbp]
\subfigure[]{\label{fig_kdvcos_U_pred_3D}
\begin{minipage}[t]{0.45\textwidth}
\centering
\includegraphics[width=\textwidth]{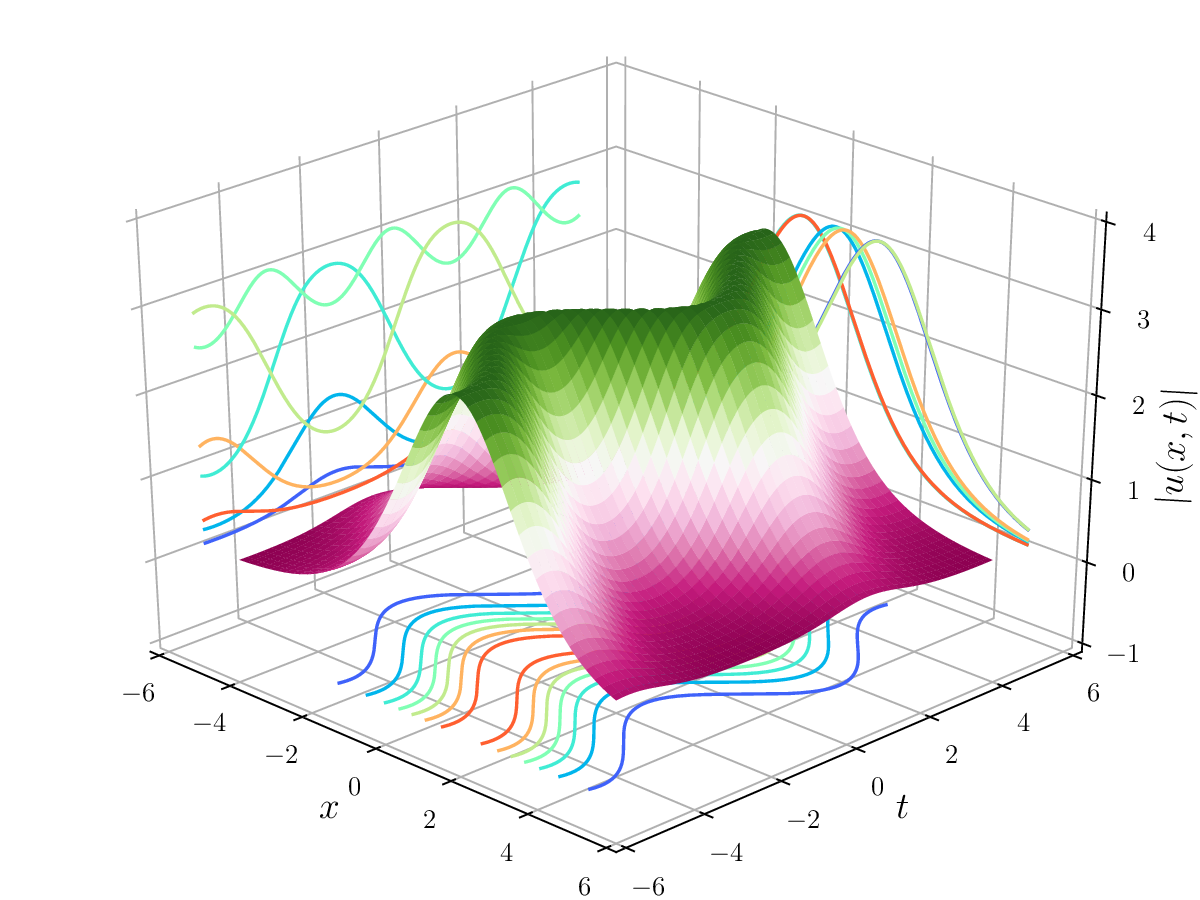}
\end{minipage}
}%
\subfigure[]{\label{fig_kdvcos_comparison_v_error2}
\begin{minipage}[t]{0.45\textwidth}
\centering
\includegraphics[width=\textwidth]{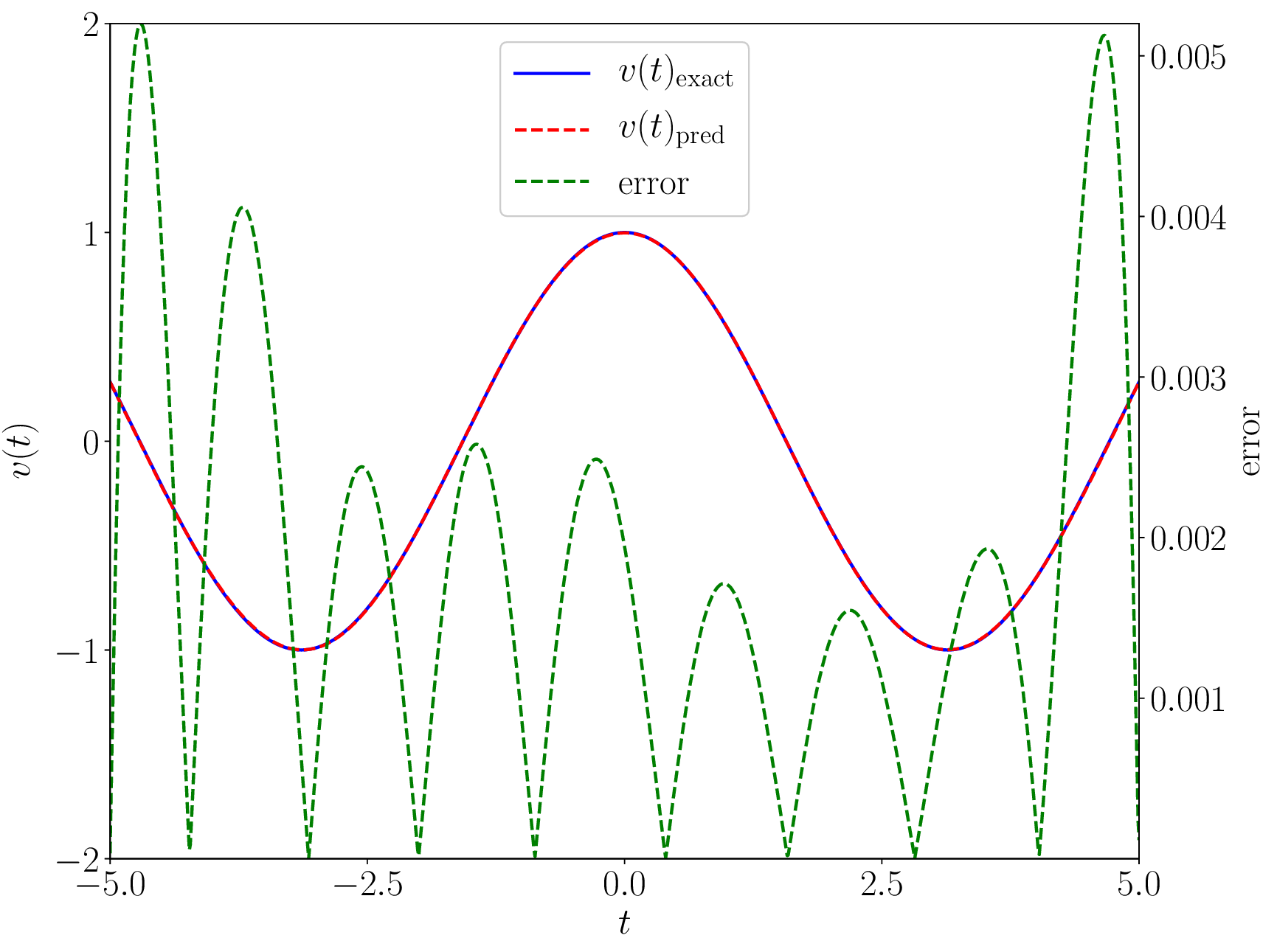}
\end{minipage}%
}%
\centering
	\caption{ (Color online) Goal function  $v(x,t)=cos(t)$: (a)3D:The predicted solution $\hat{u}$ of variable coefficient KdV equtaion within the region$[-5,5]\times[-5,5]$; (b)The data-driven variable coefficient $V(t)$ discovery of the variable coefficient KdV by R-gPINN related to solution \eqref{eq_4.1}: the comparison of the error between the learning and exact variable coefficient $V(t)$ over the interval $[-5,5]$.}
\end{figure}

The 3D plot of the predicted solution \(\hat{u}\) within the region \([-5,5] \times [-5,5]\) is shown in \ref{fig_kdvcos_U_pred_3D}. The comparison of the error between the learned and exact variable coefficient \(v(t)\) over the interval \([-1,1]\) is shown in Figure \ref{fig_kdvcos_comparison_v_error2}, with an error for \(\hat{v}(t)\) of \(2.92\times 10^{-3}\).

The Table \ref{compare-kdv-cost} compares the results of solving the inverse problem of the variable coefficient KdV equation using the R-PINN, gPINN and R-gPINN methods, with the target function \(v(t) = \cos(t)\). 
\begin{table}[htbp]
	\centering
\caption{Comparison results of R-PINN, gPINN and R-gPINN methods for the data-driven discovery of the nonlinear coefficient \( v(t) = \cos(t) \) in the variable coefficient KdV equation}
  \scalebox{0.9}{
	\begin{tabular}{|c|c|c|c|c|c|}
		\hline
		\multirow{2}{*}{Results} & \multicolumn{3}{c|}{error} & \multirow{2}{*}{$ERR_{13}$(\%)} & \multirow{2}{*}{$ERR_{23}$ (\%)} \\\cline{2-4}
                           & R-PINN &  gPINN &  R-gPINN &  &   \\                
		\hline
	\( u \) &5.61$\times 10^{-4}$& 3.45$\times 10^{-4}$ & 2.68$\times 10^{-4}$ &52.28 & 22.30  \\
		\hline
		\( v \) &2.70$\times 10^{-2}$ & 3.25$\times 10^{-3}$ & 2.92$\times 10^{-3}$ & 89.16& 9.95  \\
		\hline
	\end{tabular}}
	\label{compare-kdv-cost}
\end{table}
The results show that the error \(u\) for the gPINN method is \(3.45 \times 10^{-4}\), while the R-gPINN method has a smaller error of \(2.68 \times 10^{-4}\), representing a reduction of approximately 22.30\%. For the error \(v\), the gPINN method has an error of \(3.25 \times 10^{-3}\), whereas the R-gPINN method reduces this to \(2.92 \times 10^{-3}\), a reduction of about 9.95\%. This indicates that the R-gPINN method performs better in solving this inverse problem.

\begin{table}[h!]
\centering
\caption{Evaluation of the impact of  different residual  connection with 1, 2, and 3 hidden layers on the performance of the R-gPINN method in solving the  nonlinear coefficient  \( v(t) = \cos(t) \)  discovery  problem for the variable coefficient KdV equation.}
  \scalebox{0.9}{
\begin{tabular}{|c|c|c|c|c|c|c|c|}
\hline
\multirow{2}{*}{\diagbox{Results}{Methods}} &  \multicolumn{6}{c|}{R-gPINN} \\ \cline{2-7}
                                            & pre1 & pre2 & pre3&post1&post2&post3 \\ \hline
error$_{u}$ & 2.68$\times 10^{-4}$  & 2.62$\times 10^{-4}$  & 3.51$\times 10^{-4}$    & 1.78$\times 10^{-4}$  & 1.95$\times 10^{-4}$  & 2.45$\times 10^{-4}$ \\ \hline
error$_{v}$ & 2.92$\times 10^{-3}$  & 2.62$\times 10^{-3}$  & 3.58$\times 10^{-3}$    & 1.42$\times 10^{-3}$  & 2.28$\times 10^{-3}$  & 2.20$\times 10^{-3}$  \\ \hline
\end{tabular}}
\label{sum-kdv-cos}
\end{table}
The data in Table \ref{sum-kdv-cos} illustrates the impact of different residual connection types and the number of hidden layers on the performance of the R-gPINN method in solving the nonlinear coefficient \( v(t) = \cos(t) \) discovery problem for the variable coefficient KdV equation. Overall, the post-activation residual connection (post) outperforms the pre-activation residual connection (pre) in reducing errors. As the number of hidden layers increases, the errors for both types of residual connections also increase, suggesting that an excessive number of hidden layers may hinder model convergence in this problem. Specifically, for both error$_{u}$ and error$_{v}$, R-gPINN(post1) achieves the lowest errors, \(1.78 \times 10^{-4}\) and \(1.42 \times 10^{-3}\) respectively, indicating that the post-activation residual connection with one hidden layer yields the best performance.

\subsection{Data-driven inverse problems of the variable coefficient SG equation}

The variable coefficient SG equation is given by\cite{sg-equation}:
\begin{equation}\label{variable coefficient SGeq}
	u_{x t}+v(t) \sin (u)=0.
\end{equation}
This equation is significant in various applications such as spin-wave propagation with variable interaction strength, DNA soliton dynamics, and flux dynamics in Josephson junctions with impurities.

According to \cite{Wazwaz-2019}, the SG equation with a time-dependent coefficient is integrable and passes the Painlevé test for any analytic time-dependent coefficient. The paper also reports multiple optical kink solutions. Specifically, the single soliton solution of the variable coefficient SG equation is:

\begin{equation}\label{variable coefficient SGs}
u(x, t)=4 \arctan \left(e^{k_1 x-\omega_1(t)}\right),
\end{equation}
where \(\omega_1(t) = \int \frac{v(t)}{k_1} \, dt\) and \(k_1 \in \mathbb{R}\) is a free parameter. Setting \(k_1 = 1\), \(\omega_1(t)\) is determined by the variable coefficient \(v(t)\) in the equation.

We will investigate the target variable coefficient \(v(t)\) for both linear (\(v(t) = t\)) and nonlinear (\(v(t) = \cos(t)\)) cases. The training domain is chosen as \([-5,5] \times [-5,5]\), and 20,000 sampling points are selected within this domain using the LHS method. For the experiments, \(d\) is set to 9 and \(N_d\) to 40. We will present the results for predicting \(u\) and \(v\) using the R-gPINN method and compare the relative \(\mathbb{L}_2\) norm errors between the R-PINN, gPINN and R-gPINN methods.

\subsubsection{Data-driven discovery of linear coefficient $v(t)=t$ of the variable coefficient SG equation}

For the linear coefficient case, where \(v(t) = t\) and \(k_1 = 1\), the variable coefficient SG equation has the exact soliton solution:
\begin{equation}\label{variable coefficient SGs11}
u_{s1}(x, t) = 4 \arctan \left(e^{x - \frac{1}{2} t^2}\right).
\end{equation}

The goal function is \(v_{exact}(x,t) = t\), with the training region \([-5,5] \times [-5,5]\). The variable coefficient SG equation is considered with the following conditions:
\begin{equation}
\begin{cases}
    u_{x t} + v(x,t) \sin(u) = 0, \\
    u(x, t_0) = u_{s1}(x, -5), \\
    u(x_l, t) = u_{s1}(-5, t), \quad u(x_r, t) = u_{s1}(5, t), \quad (x, t) \in [-5,5] \times [-5,5], \\
    u(x_{in}, t_{in}) = u_{s1}(x_{in}, t_{in}), \quad (x_{in}, t_{in}) \in [-5,5] \times [-5,5], \\
    v(x, t_e) = v_{exact}(x, t_e), \quad t_e = t_0 / t_1.
\end{cases}
\end{equation}

\begin{figure}[htbp]
	\centering
		\includegraphics[width=0.8\textwidth]{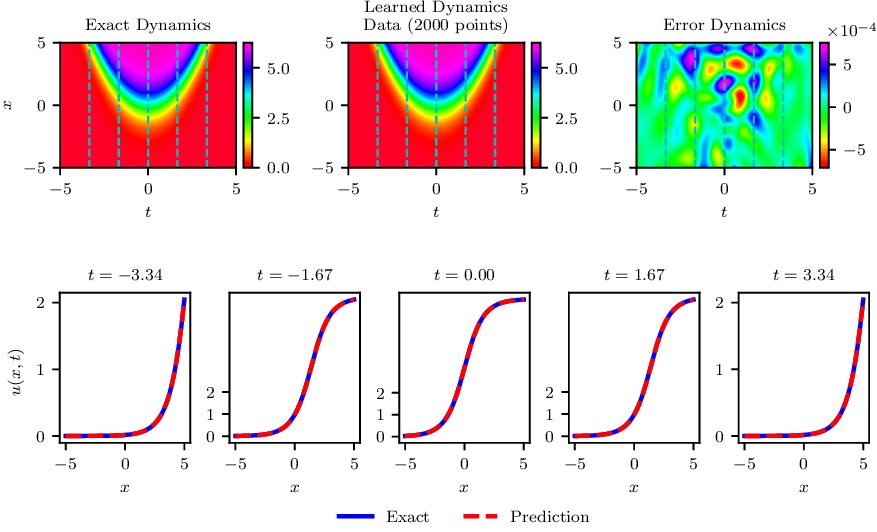}
		\caption{(Color online) Data-driven discovery of the linear coefficient \(v(t) =t\) in the variable coefficient SG equation: the upper part of the figure shows the dynamic evolution of the exact solutions, predicted solutions, and their corresponding errors; the lower part displays  the time slices at specific points in time.}
\label{fig_sgup}
\end{figure}

The linear coefficient discovery problem is solved using 3,000 iterations of Adam optimization and 6186  iterations of L-BFGS optimization, with a total training time of 647.73 seconds.
The predicted solution \(\hat{u}\) obtained through the R-gPINN method is illustrated in Figure \ref{fig_sgup}, with an error of \(7.91 \times 10^{-5}\). 

\begin{figure}[htbp]
\subfigure[]{\label{fig_sgu3d}
\begin{minipage}[t]{0.45\textwidth}
\centering
\includegraphics[width=\textwidth]{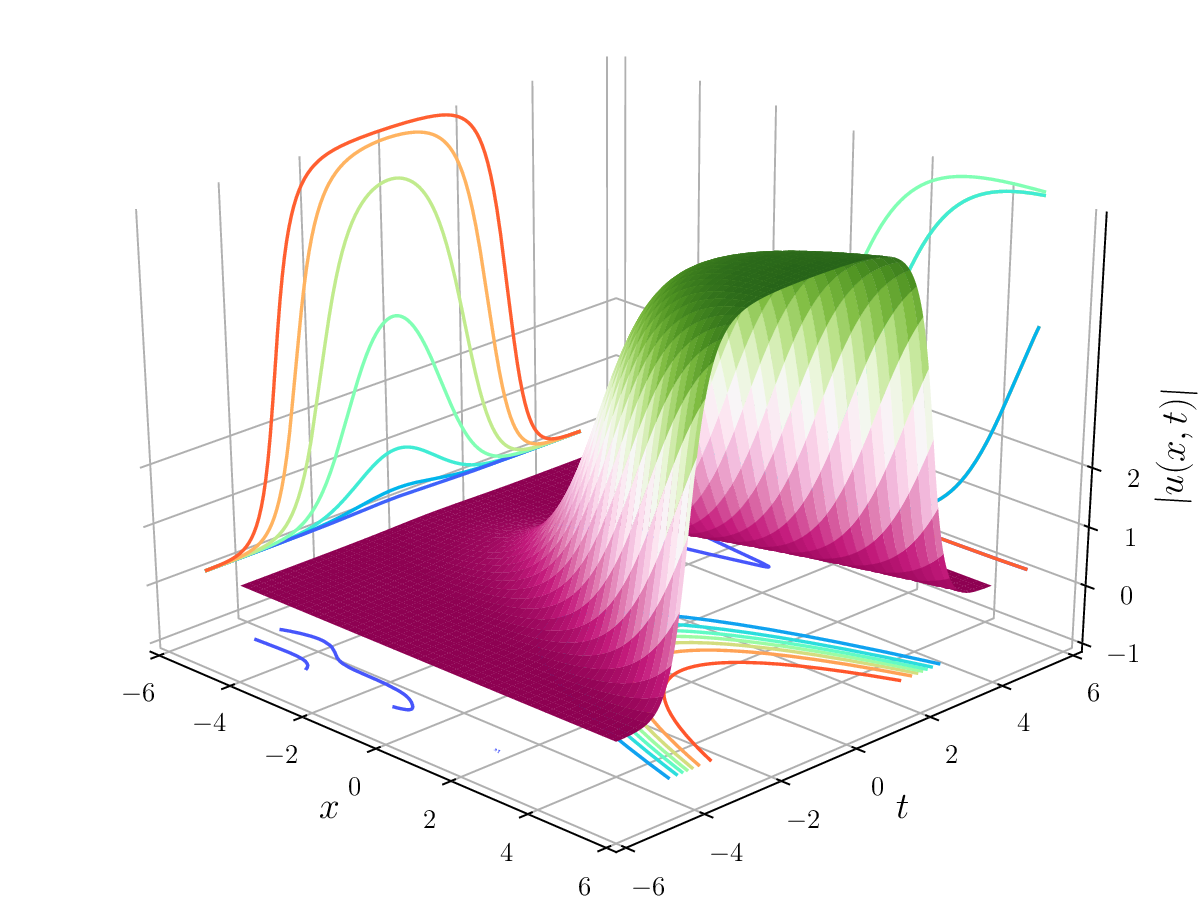}
\end{minipage}
}%
\subfigure[]{\label{sgt_comparison_v_error1}
\begin{minipage}[t]{0.45\textwidth}
\centering
\includegraphics[width=\textwidth]{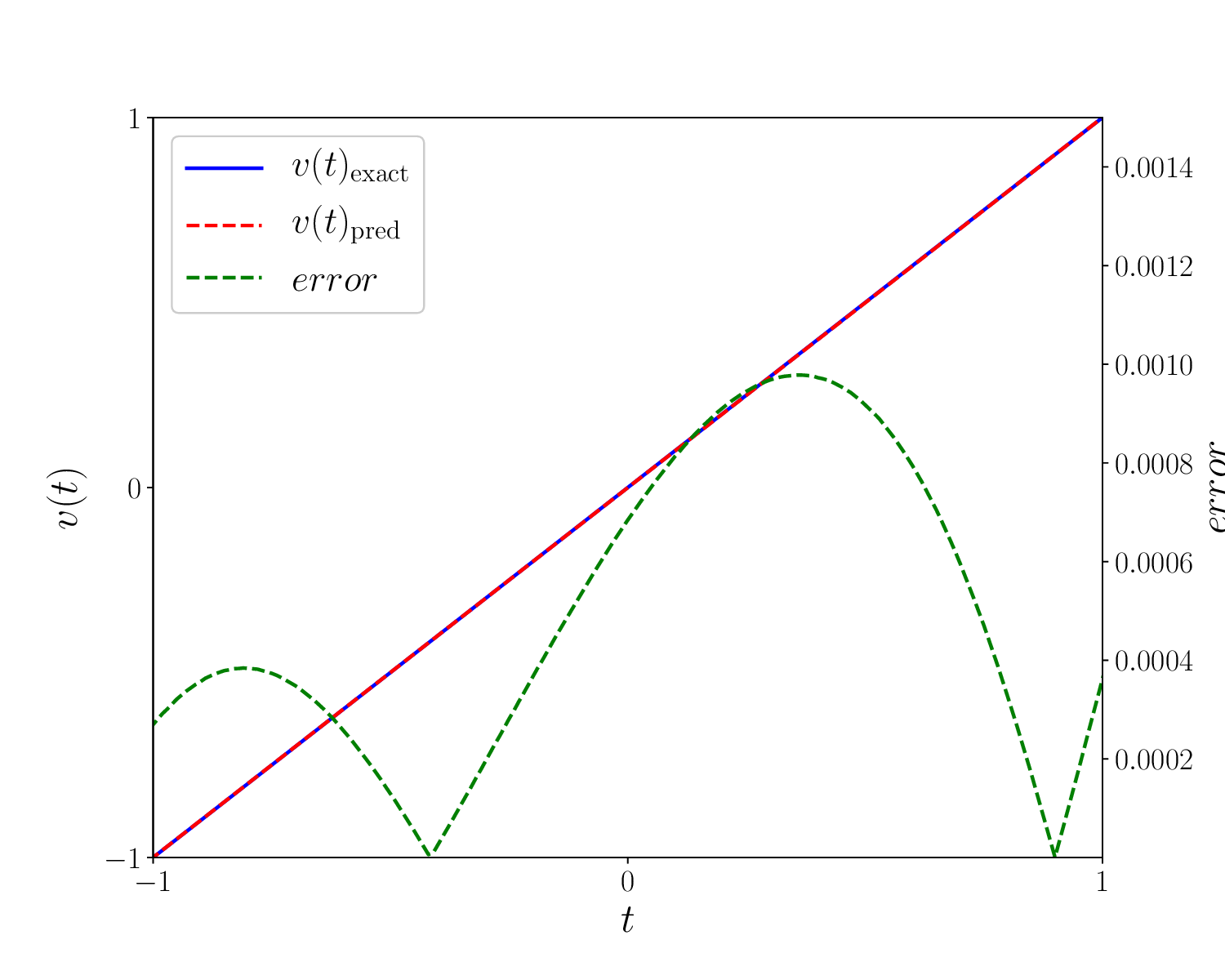 }
\end{minipage}%
}%
\centering
	\caption{ (Color online) (a)3D:The predicted solution $\hat{u}$ of variable coefficient SG equtaion within the region$[-5,5]\times[-5,5]$; (b)The predicted $\hat{v}(t)$ of variable coefficient SG equtaion with goal function  $v(x,t)=t$ over the interval $[-5,5]$}
\end{figure}

The 3D plot of predicted solution $\hat{u}$ of variable coefficient SG equtaion within the region$[-5,5]\times[-5,5]$ obtained through the R-gPINN method is depicted in Figure \ref{fig_sgu3d}, the error for $\hat{u}$ was 7.91$ \times 10^{-5}$.
 The comparison of the error between the learned and exact variable coefficient \(v(t)\) over the interval \([-5,5]\) is shown in Figure \ref{sgt_comparison_v_error1}, with an error for \(\hat{v}(t)\) of \(1.61 \times 10^{-3}\).

\begin{table}[htbp]
	\centering
\caption{ (Color online) Comparison results of R-PINN, gPINN and R-gPINN methods for the data-driven discovery of the linear coefficient \( v(t) = t \) in the variable coefficient SG equation}
  \scalebox{0.9}{
	\begin{tabular}{|c|c|c|c|c|c|}
		\hline
	\multirow{2}{*}{Results} & \multicolumn{3}{c|}{error} & \multirow{2}{*}{$ERR_{13}$(\%)} & \multirow{2}{*}{$ERR_{23}$ (\%)} \\\cline{2-4}
                           & R-PINN &  gPINN &  R-gPINN &  &   \\                
		\hline
	\( u \) & 1.84$\times 10^{-4}$ & 1.32$\times 10^{-4}$ & 7.91$\times 10^{-5}$ & 57.08 & 39.87 \\
		\hline
		\( v \) &1.75$\times 10^{-2}$ & 3.97$\times 10^{-3}$ & 1.61$\times 10^{-3}$ & 90.80& 59.45  \\
		\hline
	\end{tabular}}
	\label{compare-sg-t}
\end{table}
Table \ref{compare-sg-t} compares the R-PINN, gPINN, and R-gPINN methods for discovering the linear coefficient \( v(t) = t \) in the variable coefficient SG equation. The R-gPINN method shows the lowest errors for both \( u \) and \( v \), with values of \(7.91 \times 10^{-5}\) and \(1.61 \times 10^{-3}\), respectively. This results in a significant improvement over the gPINN method and the initial R-gPINN method, with reductions of up to 57.08\% and 90.80\%, indicating its superior performance.

\begin{table}[h!]
\centering
\caption{ (Color online) Evaluation of the impact of  different residual  connection with 1, 2, and 3 hidden layers on the performance of the R-gPINN method in solving the linear coefficient  \( v(t) = t \)  discovery  problem for the variable coefficient SG equation.}
\resizebox{\textwidth}{!}{
\scriptsize
\begin{tabular}{|c|c|c|c|c|c|c|c|}
\hline
\multirow{2}{*}{\diagbox{Results}{Methods}} &  \multicolumn{6}{c|}{R-gPINN} \\ \cline{2-7}
                                            & pre1 & pre2 & pre3&post1&post2&post3 \\ \hline
 error$_{u}$ & 7.91$\times 10^{-5}$  & 1.27$\times 10^{-4}$    & 1.39$\times 10^{-4}$ & 1.52$\times 10^{-4}$    & 1.08$\times 10^{-4}$  & 1.33$\times 10^{-4}$ \\ \hline
 error$_{v}$ & 1.61$\times 10^{-3}$   & 2.71$\times 10^{-3}$   & 6.33$\times 10^{-3}$  & 2.15$\times 10^{-3}$   & 1.98$\times 10^{-3}$  & 4.99$\times 10^{-3}$  \\ \hline
\end{tabular}
}
\label{sum-sgt}
\end{table}
Table \ref{sum-sgt} evaluates the impact of different residual connections with 1, 2, and 3 hidden layers on the R-gPINN method's performance for discovering the linear coefficient \( v(t) = t \) in the variable coefficient SG equation. The results indicate that the pre-activation residual connections generally achieve lower errors compared to post-activation connections. Specifically, the pre-activation residual connection with one hidden layer (pre1) yields the lowest errors for both \( u \) and \( v \), with \( 7.91 \times 10^{-5} \) and \( 1.61 \times 10^{-3} \), respectively. This suggests that fewer hidden layers in the pre-activation residual block contribute to better performance in this context.

\subsubsection{Data-driven discovery of nonlinear coefficient $v(t)=cos(t)$ of the variable coefficient SG equation }

When \( v(t) = \cos(t) \) and \( k_{1} = 1 \) in solution \eqref{variable coefficient SGs}, the variable coefficient SG equation has the exact soliton solution given by:
\begin{equation}\label{variable coefficient SGs2}
u_{s2}(x, t) = 4 \arctan \left(e^{x - \sin(t)}\right).
\end{equation}
The target function is \( v_{exact}(x, t) = \cos(t) \), and the training domain is \(\Omega \times \left[t_0, t_1\right] = [-5, 5] \times [-5, 5]\).

Consider the variable coefficient SG equation with the following conditions:
\begin{equation}
\begin{cases}
		u_{xt} + v(x, t) \sin(u) = 0,  \\
		u(x, t_0) = u_{s1}(x, -5), \\
		u(x_l, t) = u_{s1}(-5, t), \quad u(x_r, t) = u_{s1}(5, t), \quad (x, t) \in [-5, 5] \times [-5, 5],\\
		u(x_{in}, t_{in}) = u_{s1}(x_{in}, t_{in}), \quad (x_{in}, t_{in}) \in [-5, 5] \times [-5, 5],\\
		v(x, t_{e}) = v_{exact}(x, t_{e}), \quad t_{e} = t_{0}  /  t_{1}.
		\label{eq_sg_cost}
	\end{cases}
\end{equation}

\begin{figure}[htbp]
	\centering
		\includegraphics[width=0.8\textwidth]{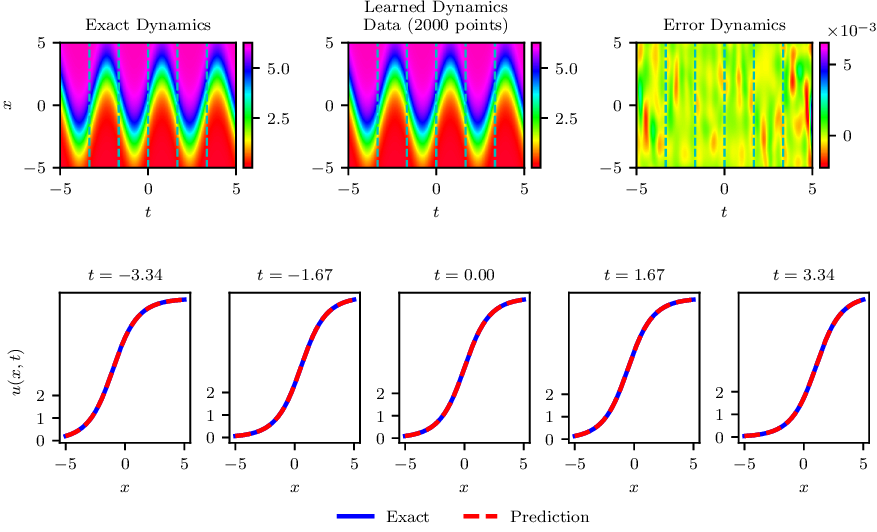}
		\caption{(Color online) Data-driven discovery of the nonlinear coefficient \(v(t) =\cos(t)\) in the variable coefficient SG equation: the upper part of the figure shows the dynamic evolution of the exact solutions, predicted solutions, and their corresponding errors; the lower part displays  the time slices at specific points in time.}
		\label{fig_pred_u_sg_cost}
\end{figure}

\begin{figure}[htbp]
\subfigure[]{\label{fig_sgcost_U_pred_3D}
\begin{minipage}[t]{0.45\textwidth}
\centering
\includegraphics[width=\textwidth]{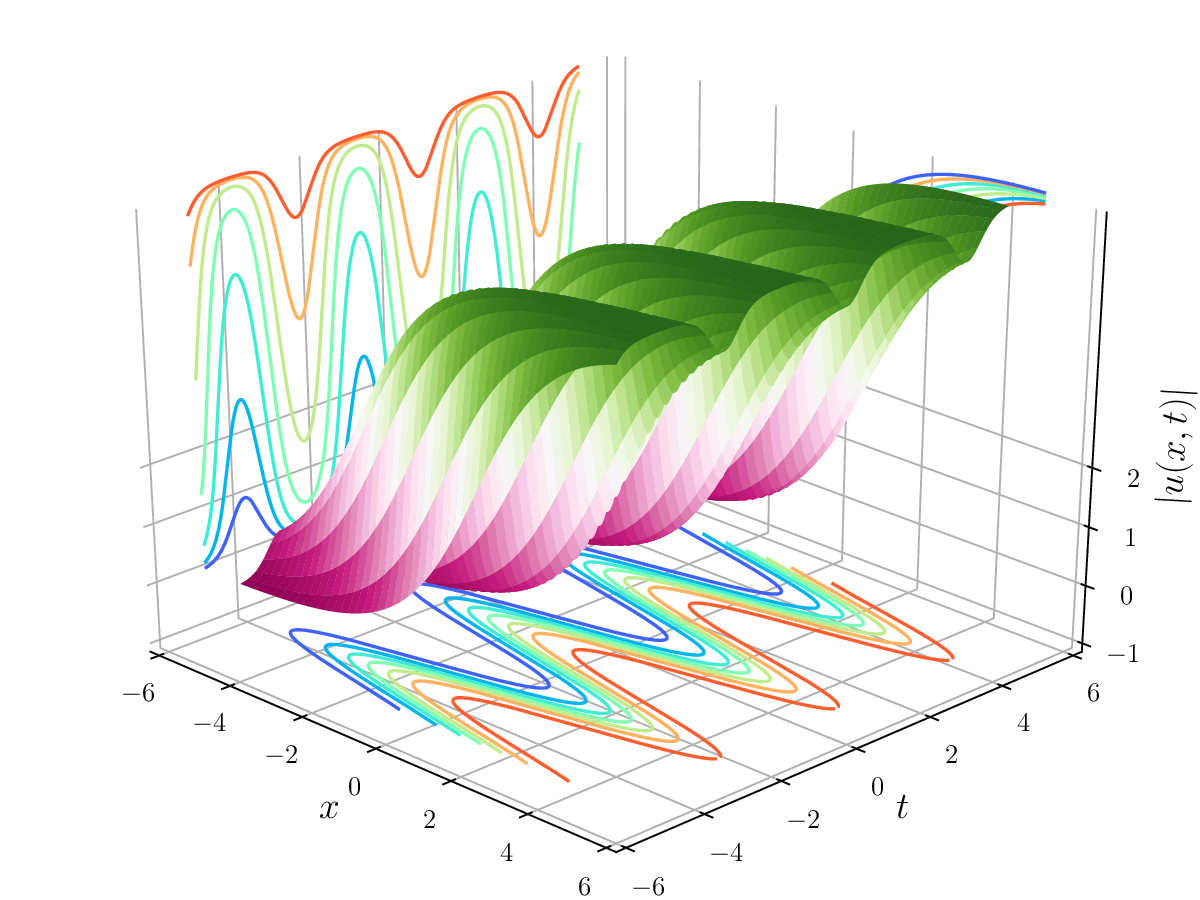}
\end{minipage}
}%
\subfigure[]{\label{fig_sgcost_comparison_v}
\begin{minipage}[t]{0.45\textwidth}
\centering
\includegraphics[width=\textwidth]{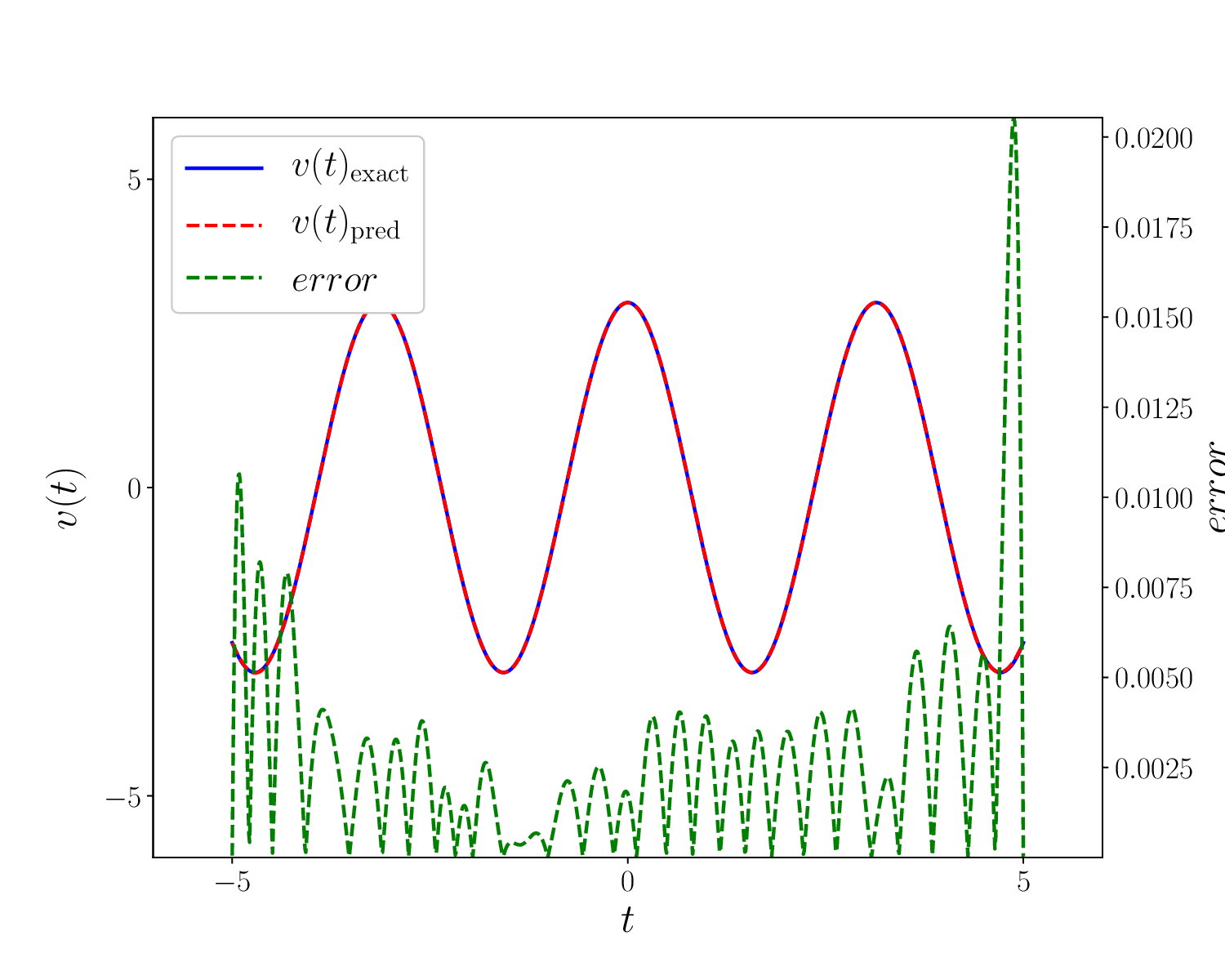}
\end{minipage}%
}%
\centering
	\caption{ (Color online) Goal function  \(v(x,t)=\cos(t)\): (a) 3D: The predicted solution \(\hat{u}\) of the variable coefficient SG equation within the region \([-5, 5] \times [-5, 5]\); (b) The predicted \(\hat{v}(t)\) of the variable coefficient SG equation over the interval \([-5, 5]\).}
\end{figure}

The predicted solution \(\hat{u}\), obtained using the R-gPINN method, is shown in Figure \ref{fig_pred_u_sg_cost}. The prediction error is \(1.39 \times 10^{-4}\), achieved after 3000 iterations of Adam optimization and 14,041 iterations of L-BFGS optimization, with a total training time of 1090.98 seconds.
The \(\hat{u}\) obtained through the R-gPINN method  is depicted in Figure \ref{fig_sgcost_U_pred_3D}.  The error between the predicted  \(\hat{v}(x,t)\) and exact variable coefficient \(v(x,t)\) over the interval \([-5, 5]\) is \(1.91 \times 10^{-3}\). Setting \( v(t) = {v}(x_{i}, t) \), the comparison of the error between the predicted  \(\hat{v}(t)\) and exact variable coefficient \(v(t)\) over the interval \([-5, 5]\) is illustrated in Figure \ref{fig_sgcost_comparison_v}.

\begin{table}[htbp]
	\centering
\caption{ Comparison results of R-PINN, gPINN and R-gPINN methods for the data-driven discovery of the nonlinear coefficient \( v(t) = \cos(t) \) in the variable coefficient SG equation}
  \scalebox{0.9}{
	\begin{tabular}{|c|c|c|c|c|c|}
		\hline
	\multirow{2}{*}{Results} & \multicolumn{3}{c|}{error} & \multirow{2}{*}{$ERR_{13}$(\%)} & \multirow{2}{*}{$ERR_{23}$ (\%)} \\\cline{2-4}
                           & R-PINN &  gPINN &  R-gPINN &  &   \\                
		\hline
		\( u \) &3.46$\times 10^{-4}$ & \(2.31 \times 10^{-4}\) & \(1.38 \times 10^{-4}\) & 59.91 & 39.98 \\
		\hline
		\( v\) &1.49$\times 10^{-2}$ & \(3.01 \times 10^{-3}\) & \(1.91 \times 10^{-3}\) & 87.15 & 36.48  \\
		\hline
	\end{tabular}}
	\label{compare-sg-cost}
\end{table}
Table \ref{compare-sg-cost} presents the comparison results of the R-PINN, gPINN, and R-gPINN methods for the data-driven discovery of the nonlinear coefficient \( v(t) = \cos(t) \) in the variable coefficient SG equation. The error values indicate that the R-gPINN method achieved the lowest errors for both variables \( u \) and \( v \), with values of \(1.38 \times 10^{-4}\) and \(1.91 \times 10^{-3}\) respectively, demonstrating superior performance. In contrast, the R-PINN method had the highest errors, suggesting that it is less effective for this problem compared to the other two methods. Specifically, the R-gPINN method reduced the errors by 59.91\% and 87.15\% for \( u \) and \( v \) respectively, compared to the R-PINN method, and by 39.98\% and 36.48\% respectively, compared to the gPINN method. These results indicate that the R-gPINN method provides better accuracy and robustness in handling inverse problems involving nonlinear coefficients.

\begin{table}[h!]
\centering
\caption{Evaluation of the impact of  different residual  connection with 1, 2, and 3 hidden layers on the performance of the R-gPINN method in solving the  nonlinear coefficient  \( v(t) = \cos(t) \) discovery  problem for the variable coefficient SG equation.}
  \scalebox{0.9}{
\begin{tabular}{|c|c|c|c|c|c|c|c|}
\hline
\multirow{2}{*}{\diagbox{Results}{Methods}} &  \multicolumn{6}{c|}{R-gPINN} \\ \cline{2-7}
                                            & pre1 & pre2 & pre3&post1&post2&post3 \\ \hline
error$_{u}$ & 1.39$\times 10^{-4}$ & 2.31$\times 10^{-4}$ & 3.63$\times 10^{-4}$  &  2.10$\times 10^{-4}$    & 1.38$\times 10^{-4}$  &  2.10$\times 10^{-4}$ \\ \hline
error$_{v}$ & 1.91$\times 10^{-3}$ & 2.71$\times 10^{-3}$  & 4.92$\times 10^{-3}$ & 3.09$\times 10^{-3}$    & 1.87$\times 10^{-3}$  & 2.82$\times 10^{-3}$  \\ \hline
\end{tabular}}
\label{sum-sgcost}
\end{table}
Table \ref{sum-sgcost} presents the impact of different residual connection types and the number of hidden layers on the performance of the R-gPINN method in solving the nonlinear coefficient \( v(t) = \cos(t) \) discovery problem for the variable coefficient SG equation. The results show that with two hidden layers, R-gPINN(post2) achieved the optimal error$_{u}$ and error$_{v}$, with values of \(1.38 \times 10^{-4}\) and \(1.87 \times 10^{-3}\) respectively. In contrast, as the number of hidden layers increases, the errors generally tend to rise, particularly in the case of three hidden layers, where the error$_{v}$ significantly increases to \(4.92 \times 10^{-3}\). This suggests that an excessive number of hidden layers can lead to a decline in model performance, while the post-activation residual connection with two hidden layers (post2) provides the best balance for this problem.

\subsection{Data-driven inverse problems of the variable coefficient KP equation}

The KP equation, discovered by Kadomtsev and Petviashvili while studying nonlinear waves in weakly dispersed media\cite{KP-1970}, describes the propagation of weakly nonlinear, weakly dispersed waves in one direction with small perturbations in others. It has broad applications in fluid mechanics, plasma physics, and gas dynamics, serving as a model for (2+1)-dimensional shallow water waves and ion acoustic waves. While the standard KP equation has been widely studied, research on the variable coefficient KP equation is limited. However, the variable coefficient KP equation offers a more accurate representation of surface waves, especially in scenarios where waves traverse through regions with varying width, depth, and density, such as entering the sea or ocean through a canyon. This makes it more effective at describing real-world wave conditions than the constant coefficient KP equation \cite{david-1987,david-1989,gungor-2002}. In \cite{Miao1}, a specific form of a  generalized variable coefficient  KP equation is  given as follows:
\begin{equation}\label{vckp-eq}
(u_t+f(t) u u_x+g(t) u_{x x x}+l(t)u)_{x}+m(t) u_{yy}=0,
\end{equation}
here $u = u(x, y, t)$, $x, y$ are space variables, and $t$ is time variable.

Considering the following constraints:
\begin{equation}
\begin{split}
 g(t)=\gamma f(t) e^{-\int l(t) d t}, \\
 m(t)=\rho f(t) e^{-\int l(t) d t},
\end{split}
\end{equation}
where $\gamma$ and $\rho$ are arbitrary parameters. 

Let $\gamma$= $\rho$ = 1,  $f(t)=\sin (t), l(t)=\frac{1}{10}$, then a exact solution of \eqref{vckp-eq} is obtained as follows:
$$
u_{kp1}=\frac{12 e^{-\frac{t}{10}+x+\sqrt{3} y+\frac{40}{101}  e^{-\frac{t}{10}}[10 \cos (t)+\sin (t)]}}{\left(1+e^{x+\sqrt{3} y+\frac{40}{101}  e^{-\frac{t}{10}}[10 \cos (t)+\sin (t)]}\right)^2} .
$$
The goal function $v_{exact}(x,y,t)=g(t)=m(t)= e^{-\frac{t}{10}}sin(t)$ and the training region is taken as \( [-6,6] \times [-6,6]\times [-6,6]\).

\begin{figure}[htbp]
    \centering
    \includegraphics[width=\textwidth]{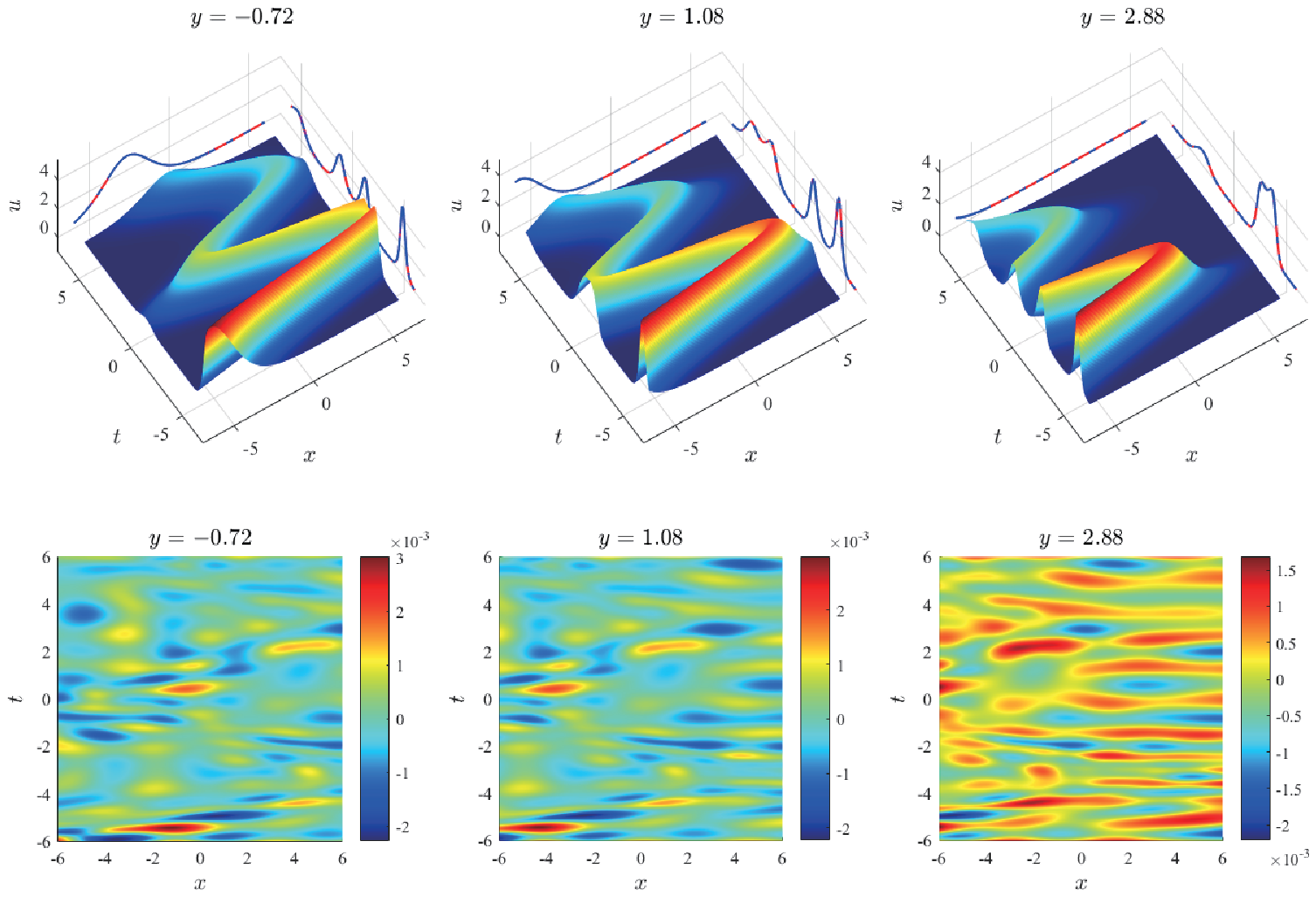}
\caption{(Color online) Data-driven discovery of the nonlinear coefficient \(v(t) =\cos(t)\) in the variable coefficient KP equation: 
the upper part of the figure displays the 3D predicted solution at these three cross-sections. The red dashed lines represent the predicted solution, while the blue lines represent the true solution;  the lower part displays the errors between the predicted and true solutions at the same cross-sections. }
\label{fig_KP3d}
\end{figure}

Consider the  variable coefficient KP equation with initial and boundary conditions:
\begin{equation}\label{ib-vckp}
\begin{cases}
	u_xt+sin(t)(u u_xx+u_x u_x)+V(t) u_{x x x x}+\frac{1}{10}u_{x}+V(t) u_{yy}=0,  \\
		u(x,y,t_0) = u_{kp1}(x,y,-6), \\
		u(x_l,y,t) = u_{kp1}(-6,y,t), \\
		u(x_r,y,t) = u_{kp1}(6,y,t), \\
		u(x,y_l,t) = u_{kp1}(x,-6,t), \\
		u(x,y_r,t) = u_{kp1}(x,6,t), \\
		u(x_{in},y_{in},t_{in})=u_{kp1}(x_{in},y_{in},t_{in}), (x_{in},y_{in},t_{in}) \in [-6,6] \times [-6,6]\times [-6,6],\\
		v(x,y,t_{e})=v_{exact}(x,y,t_{e}),t_{e}=t_{0}/t_{1}.
	\end{cases}
\end{equation}
MATLAB is used to discretize the domain into \(N_x=N_y=N_t=101\) equidistant points in the spatial and temporal directions, respectively. From this discretized domain, we select \(N_{u}=0\) points corresponding to the initial and boundary conditions, as well as \(N_{u_{in}}=5000\) discrete points within the domain. We employ a 5-layer feedforward neural network with 30 neurons per layer, incorporating pre-activated residual units with a single hidden layer within each residual block, to solve nonlinear coefficient discovery problems for the variable coefficient KP equation. After  3000  iterations of Adam and 7108 iterations of L-BFGS optimization, with a total training time of 1131.26 seconds.  We obtain the predict solution \(\hat{u}\) and predict function \(\hat{v}(t)\). The prediction error for \(\hat{u}\) is \( 5.09 \times 10^{-4}\), and the error for \(\hat{v}(t)\) is \(1.83 \times 10^{-3}\). The predicted solution \(\hat{u}\) obtained through the R-gPINN method is shown in Figure \ref{fig_KP3d}.  

\begin{figure}[htbp]
	\centering
	\includegraphics[width=0.45\textwidth]{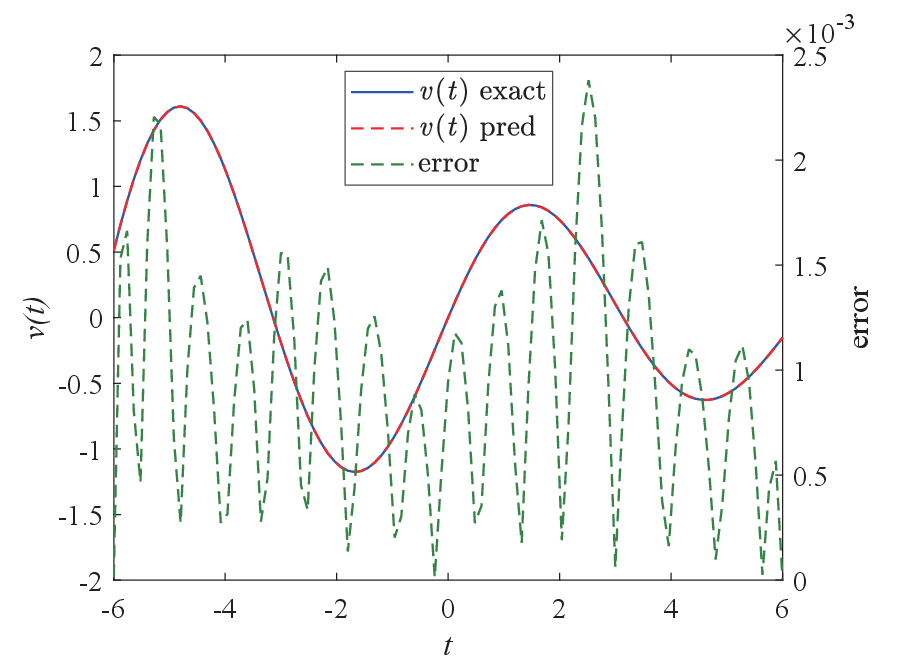}
	\caption{ (Color online) The data-driven variable coefficient $V(t)$ discovery of the variable coefficient KP equation by R-gPINN related to solution \eqref{eq_4.1}: the comparison of the error between the learning and exact variable coefficient $V(t)$ over the interval $[-6,6]$.}
	\label{fig_comparison_kp_v_error}
\end{figure}

 Setting \( v(t) = {v}(x_{i},y_{i}, t) \), the comparison of the error between the predicted  \(\hat{v}(t)\) and exact variable coefficient \(v(t)\) over the interval \([-6, 6]\) is illustrated in Figure \ref{fig_comparison_kp_v_error}.

Table \ref{compare-kp} present the comparison results of the R-PINN, gPINN, and R-gPINN methods for data-driven function discovery of variable coefficient KP equation. Furthermore, Table \ref{sum-kp} evaluate the impact of different residual connections with 1, 2, and 3 hidden layers on the performance of the R-gPINN method in solving the data-driven function discovery of variable coefficient KP equation.
\begin{table}[htbp]
	\centering
\caption{Comparison results of R-PINN, gPINN and R-gPINN methods for the data-driven discovery of the nonlinear coefficient \( v(t) = \cos(t) \) in the variable coefficient KP equation}
  \scalebox{0.9}{
	\begin{tabular}{|c|c|c|c|c|c|}
		\hline
	\multirow{2}{*}{Results} & \multicolumn{3}{c|}{error} & \multirow{2}{*}{$ERR_{13}$(\%)} & \multirow{2}{*}{$ERR_{23}$ (\%)} \\\cline{2-4}
                           & R-PINN &  gPINN &  R-gPINN &  &   \\                
		\hline
	\( u \) & 8.14$\times 10^{-4}$    & 6.12$\times 10^{-4}$ & 5.09$\times 10^{-4}$ &37.44& 16.77  \\
		\hline
		\( v \) &1.69$\times 10^{-1}$     & 2.56$\times 10^{-3}$ & 1.83$\times 10^{-3}$ &98.91& 28.23  \\
		\hline
	\end{tabular} }
	\label{compare-kp}
\end{table}
The table shows that the R-gPINN method outperforms both R-PINN and gPINN in predicting the nonlinear coefficient \( v(t) = \cos(t) \) in the variable coefficient KP equation. Specifically, the R-gPINN method achieves the lowest prediction errors for both \( u \) and \( v(t) \), with reductions of 37.44\% and 98.91\% in error compared to R-PINN for \( u \) and \( v(t) \), respectively. Additionally, R-gPINN also demonstrates a notable improvement over gPINN, reducing the prediction errors by 16.77\% for \( u \) and 28.23\% for \( v(t) \). This analysis highlights the superior accuracy of the R-gPINN method for this problem.

\begin{table}[htbp]
\centering
\caption{Evaluation of the impact of  different residual  connection with 1, 2, and 3 hidden layers on the performance of the R-gPINN method in solving the  nonlinear coefficient  \( v(t) = \cos(t) \) discovery  problem for the variable coefficient KP equation.}
  \scalebox{0.9}{
\begin{tabular}{|c|c|c|c|c|}
\hline
\multirow{2}{*}{\diagbox{Results}{Methods}} &  \multicolumn{4}{c|}{R-gPINN} \\ \cline{2-5}
                                            & pre1 & pre2 & post1&post2 \\ \hline
error$_{u}$ & 5.09$\times 10^{-4}$   & 4.75$\times 10^{-4}$   & 5.99$\times 10^{-4}$  & 5.06$\times 10^{-4}$  \\ \hline
error$_{v}$ & 1.83$\times 10^{-3}$   & 2.23$\times 10^{-3}$  & 2.46$\times 10^{-3}$  & 2.98$\times 10^{-3}$ \\ \hline
\end{tabular}}
\label{sum-kp}
\end{table}
This table evaluates the performance of the R-gPINN method with different residual connections and varying numbers of hidden layers in solving the nonlinear coefficient \( v(t) = \cos(t) \) discovery problem for the variable coefficient KP equation. It shows that the pre-activated residual connection generally leads to lower errors for both \( u \) and \( v(t) \) compared to the post-activated connection. Specifically, the R-gPINN(pre2) configuration achieves the lowest error for \( u \) at \( 4.75 \times 10^{-4} \) and performs better for \( v(t) \) with an error of \( 1.83 \times 10^{-3} \). These results indicate that pre-activated connections, particularly with two hidden layers, yield superior performance in this problem.

\section{Conclusions and discussions}

In this paper, a novel method named  R-gPINN  is proposed for improve predict accuracy in solve forward-inverse problems of variable coefficient PDEs. The method combines the benefits of variable coefficient gradient-enhanced effects with the identity mapping ability of Residual units.  By extending the traditional PINN framework, on the one hand, R-gPINN incorporates residual unit to mitigate the gradient vanishing and network degradation, uniform linear and nonlinear coefficient problems.  On the another hand, R-gPINN integrates the gradient of variable coefficients into the loss function to enhance the physical constraints of the neural network, guiding the prediction function to progressively align with the objective function. 

Our numerical experiments, which include  solving variable coefficient Burgers equation,  variable coefficient KdV equation, variable coefficient SG equation, and high-dimensional variable coefficient KP equation. The results indicate that the R-gPINN method demonstrates significant advantages over the baseline methods, gPINN and R-PINN, in terms of error reduction rates for both the predicted solution \(\hat{u}\) and the predicted function \(\hat{v}(t)\). For the predicted solution \(\hat{u}\) , the error reduction rate of the R-gPINN method typically exceeds 20\%, reaching over 50\% in some examples. In the case of the predicted function \(\hat{v}(t)\), the error reduction rate is even more pronounced, generally surpassing 40\% and reaching up to almost 100\% in certain tests. In conclusion, the numerical experiments results demonstrate that the R-gPINN method significantly improve the generalization ability of the network in the forward-inverse problems of the variable coefficient PDEs.

In this paper, when solving inverse problems for partial differential equations, the measurement data are selected by randomly choosing a small amount of data from the entire domain. However, in certain practical scenarios, measurement data cannot be arbitrarily selected across the entire domain. Therefore, to validate the generality and effectiveness of our method, we randomly selected a small subregion of the entire spatiotemporal domain as the measurement data. For example, we randomly chose $\Omega_0$ = $\frac{1}{3}\Omega$ and $\left[t_01, t_11\right] =  \frac{1}{3}\left[t_0, t_1\right]$. Except for the data selection method, all other parameters remained consistent with those in the paper. We used the variable-coefficient Burgers equation and the variable-coefficient KP equation as examples. For the variable-coefficient Burgers equation with the target function $v(t) = t$, we obtained the predicted solution \(\hat{u}\)  and the predicted function \(\hat{v}\) . The prediction error for \(\hat{u}\)  was  \( 2.06 \times 10^{-3}\), and the error for \(\hat{v}(t)\)  was \( 6.65 \times 10^{-3}\) . For the variable-coefficient KP equation with the target function $v(t) = cos(t)$, we obtained the predicted solution \(\hat{u}\)  and the predicted function\(\hat{v}(t)\) . The prediction error for \(\hat{u}\)  was  \( 9.80 \times 10^{-3}\), and the error for \(\hat{v}\)  was   \( 2.22 \times 10^{-2}\). The results demonstrate that the proposed method remains effective when using the measurement data suggested herein. However, under the condition of a fixed total amount of measurement data, selecting points in a dispersed manner leads to stronger generalization capabilities of the trained network compared to selecting points in a concentrated manner.

The analysis of the data in this study also reveal that pre-activation residual unit of R-gPINN generally demonstrate superior performance compared to post-activation residual unit, achieving notable error reductions in both predict solutions and predict function. Nonetheless, performance variability across different test cases suggests that the optimal way may vary depending on the specific problem  underscoring the importance of selecting the appropriate residual unit configuration for each problem.
 Since both network parameters and the type of residual networks significantly impact the experimental results, our future work will focus on investigating the effects of incorporating techniques such as adaptive sampling, adaptive activation, and different types of residual units on solving forward-inverse problems of variable coefficient equations. Additionally, we will explore the integration of further physical information to optimize the performance in solving variable coefficient PDEs.

\end{document}